\documentclass{aa}

\usepackage{graphicx}
\usepackage{txfonts}

\def\vel{{\rm v}}

\begin{document}

\title{Kinematics of the Old Stellar Population at the Galactic Centre  \thanks{Based on observations at the Very Large Telescope (VLT) of the European Southern Observatory (ESO), Cerro Paranal, Chile}}

\author{S. Trippe \inst{1} \and S. Gillessen \inst{1} \and O.E. Gerhard \inst{1}  \and H. Bartko \inst{1} \and T.K. Fritz \inst{1}  \and H.L. Maness \inst{3} \and F. Eisenhauer \inst{1} \and F. Martins \inst{2} \and \\ T. Ott \inst{1} \and K. Dodds-Eden  \inst{1} \and R. Genzel \inst{1,3}}

\offprints{S. Trippe, \email{trippe@mpe.mpg.de}}

\institute{Max-Planck-Institut f\"ur extraterrestrische Physik, Giessenbachstrasse 1, D-85748 Garching, Germany \and GRAAL-CNRS, Universit\'e Montpellier II, Place Eug\`ene Bataillon, F-34090 Montpellier, France \and University of California, 94720 Berkeley, California, USA}

\date{Received May 13, 2008/ Accepted October 6, 2008}

\abstract
{}
{
We aim at a detailed description of the kinematic properties of the old, (several Gyrs) late-type CO-absorption star population among the Galactic centre (GC) cluster stars. This cluster is composed of a central supermassive black hole (Sgr~A*) and a self-gravitating system of stars. Understanding its kinematics thus offers the opportunity to understand the dynamical interaction between a central point mass and the surrounding stars in general, especially in view of understanding other galactic nuclei.
}
{
We applied AO-assisted, near-infrared imaging and integral-field spectroscopy using the instruments NAOS/CONICA and SINFONI at the VLT. We obtained proper motions for 5445 stars, 3D velocities for 664 stars, and acceleration limits (in the sky plane) for 750 stars. Global kinematic properties were analysed using velocity and velocity dispersion distributions, phase-space maps, two-point correlation functions, and the Jeans equation.
}
{
We detect for the first time significant cluster rotation in the sense of the general Galactic rotation in proper motions. Out of the 3D velocity dispersion, we derive an improved statistical parallax for the GC of $R_0=8.07\pm0.32_{\rm stat}\pm0.13_{\rm sys}~{\rm kpc}$. The distribution of 3D stellar speeds can be approximated by local Maxwellian distributions. Kinematic modelling provides deprojected 3D kinematic parameters, including the mass profile of the cluster. We find an upper limit of 4\% for the amplitude of fluctuations in the phase-space distribution of the cluster stars compared to a uniform, spherical model cluster. Using upper limits on accelerations, we constrain the minimum line-of-sight distances from the plane of Sgr~A* of five stars located within the innermost few (projected) arcsec. The stars within 0.7'' radius from the star group IRS13E do not co-move with this group, making it unlikely that IRS13E is the core of a substantial star cluster. Overall, the GC late-type cluster is described well as a uniform, isotropic, rotating, dynamically relaxed, phase-mixed system.
}
{}

\keywords{Galaxy: centre -- Galaxy: kinematics and dynamics -- Stars: kinematics -- Infrared: stars}

\maketitle

\section{Introduction}

\begin{figure*}

\begin{center}

\includegraphics[width=17cm,angle=-90]{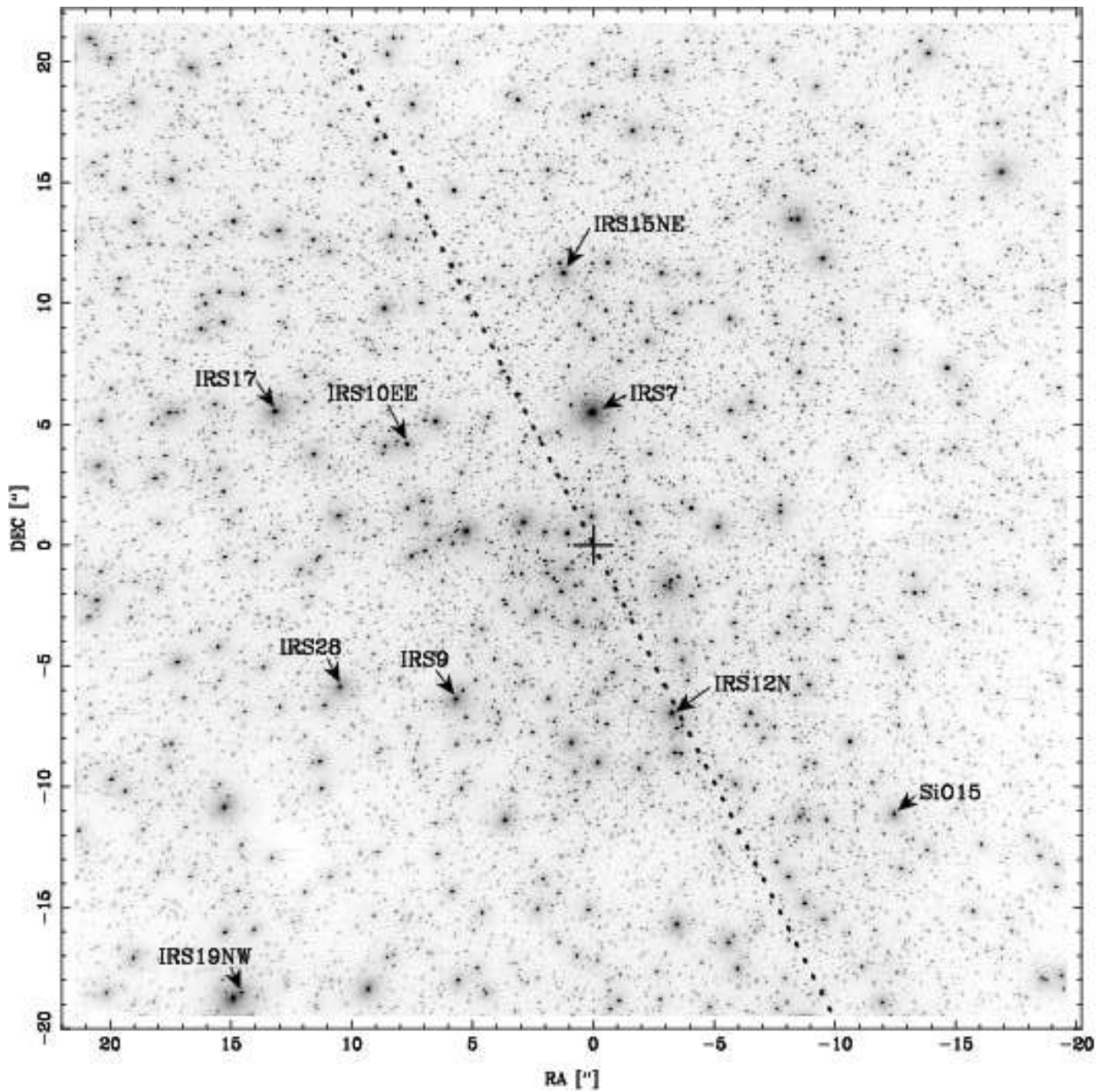}

\end{center}

\caption{K-band mosaic of the Galactic centre cluster obtained in March 2007; coordinates are R.A. and DEC in arcsec relative to Sgr~A*. This image covers a FOV of $\approx$1.5$\times$1.5~pc. The contrast was enhanced by applying unsharp masking. The position of Sgr~A* is marked with a cross in the image centre. The dotted line crossing the image indicates the Galactic plane. Stars with names are SiO maser stars used for defining an astrometric reference frame.}

\label{field}

\end{figure*}

The dynamical properties of the Galactic centre (GC) star cluster, which hosts the radio source and supermassive black hole (SMBH) Sagittarius~A* (Sgr~A*), have been subject to intensive research for about two decades. Due to strong interstellar extinction ($A_{V}\approx 30$) the GC stars can be observed only in the infrared; most of the work in this field is based on near-infrared (NIR) data ranging from H to L bands (1.5 -- 4~$\mu$m; see Fig.~\ref{field} for an example).

Initially, the central question of this research was whether the GC indeed hosts a central SMBH (e.g. Lynden-Bell \& Rees \cite{lyndenbell1971}), which was discussed even before the discovery of the radio point source Sgr~A* (Balick \& Brown \cite{balick1974}). Based on statistical arguments using the observed velocity dispersions, it was possible to show in the late 1980s and early 1990s that a central pointlike mass of a few million solar masses was present. Additionally, increasingly better estimates of the distance to the GC became possible (McGinn et al. \cite{mcginn1989}; Krabbe et al. \cite{krabbe1995}; Eckart \& Genzel \cite{eckart1997}; Ghez et al. \cite{ghez1998}; Genzel et al. \cite{genzel1996, genzel1997, genzel2000}).

With improved data quality, especially due to the establishment of speckle imaging and adaptive optics (AO) assisted imaging and spectroscopy, as well as longer observation time lines, more direct tests of the central mass were executed. These efforts led to the observation of Keplerian star orbits in the immediate vicinity ($\approx$0.5'' or $\approx$4000~AU) of Sgr~A* which allowed a direct geometric determination of the mass $M_{\bullet}$ of and the distance $R_0$ to the central SMBH (Sch\"odel et al. \cite{schoedel2002,schoedel2003}; Ghez et al. \cite{ghez2003, ghez2005, ghez2008}; Eisenhauer et al. \cite{eisenhauer2003a, eisenhauer2005}). Throughout this paper, we adopt a canonical distance $R_0=8$~kpc and a distance-scaled mass $M_{\bullet}=(4.1\pm0.4)\times10^{6}\times(R_0/{\rm 8 kpc})^{2.3}~\rm M_{\odot}$. The scaled mass is taken from Eisenhauer et al. (\cite{eisenhauer2005}). Initially, Eisenhauer et al. (\cite{eisenhauer2005}) derived a distance $R_0=7.62\pm0.32$~kpc. A recent re-analysis of the data by Gillessen et al. (\cite{gillessen2008}) showed that systematic errors are present in the data which had been neglected in earlier works (also Ghez et al. \cite{ghez2008}). In total, the uncertainty on $R_0$ is $\approx$0.4~kpc (the 0.32~kpc quoted by Eisenhauer et al. \cite{eisenhauer2005} are the statistical error). We therefore decided to stick to the canonical distance of 8.0~kpc. For  this GC distance the image scale is $\rm 1~arcsec\simeq39~mpc\simeq8000~AU$ in position and 1~mas/yr~$\simeq$~38~km/s in velocity.

In addition to these advances, a dynamically complex structure of the central cluster on scales of $\approx$1--10'' emerged. It was possible to show that the cluster is composed of two main populations: (1) a population of dynamically relaxed, evolved, old (several Gyr), spherically distributed late-type CO absorption line stars, and (2) a relatively small population of young ($\approx$6~Myr) OB- and Wolf-Rayet-stars, located in the central arcsecond and in two disks centered on Sgr~A* (Genzel et al. \cite{genzel2003}; Paumard et al. \cite{paumard2006}; Maness et al. \cite{maness2007}).

As most, if not all, nearby galactic systems contain central SMBHs (e.g. Ferrarese \& Ford \cite{ferrarese2005}, and references therein), analyzing the Galactic centre system enhances the understanding of galaxy cores in general. Stellar dynamics is an important tool for the analysis of the central masses (see, e.g., Kormendy \& Bender \cite{kormendy1999}, Bender et al. \cite{bender2005} for the case of M31). In contrast to other galactic nuclei, the centre of the Milky Way can be observed on physical scales small enough to observe accelerations of individual stars within a reasonable (few years) amount of time. Thus the GC is a unique laboratory for studying the dynamical interaction of a SMBH with its immediate stellar environment.

In this article we focus on the properties of the population of evolved late type, CO absorption stars. We present the as yet most precise kinematical analysis of the central star cluster. This work is based on proper motions and radial velocities extracted from diffraction limited imaging and spectroscopy data obtained from 2002 to 2007.

This paper is organized as follows. In Section 2, we summarize the data acquisition and reduction. Section 3 describes the extraction of stellar positions and proper motions from imaging data. Section 4 gives an overview on the collection of line-of-sight velocities from integral-field spectroscopy data. In Section 5, we present our findings and discuss them. Section 6 summarizes our results and conclusions.

\section{Observations and data processing}

\subsection{Imaging}

This work is based on observations with the 8-m-UT4 (Yepun) of the ESO-VLT on Cerro Paranal, Chile. For obtaining imaging data we used the detector system NAOS/CONICA (NACO for short) consisting of the AO system NAOS (Rousset et al. \cite{rousset2003}) and the 1024$\times$1024-pixel NIR camera CONICA (Hartung et al. \cite{hartung2003}).

\subsubsection{Data reduction}

We obtained 10 data sets in H and K bands with a pixel scale of 27 mas/pixel (large scale) covering 6 epochs (May 2002, May 2003, June 2004, May 2005, April 2006, March 2007). In this mode each image covers a field of view (FOV) of 28$\times$28''. During each observation the camera pointing was shifted such that the FOV is $\approx$40''$\times$40'' for a typical data set, centered on Sgr~A*.

We executed a much larger number of observations using a smaller pixel scale of 13 mas/pixel (small scale), thus resulting in an image FOV of 14''$\times$14'' and, again using shifted pointings, typical observation FOVs of 20''$\times$20''. In total we obtained 42 H and K band image sets, 5 to 10 per year with a roughly monthly sampling.

To all images we applied sky-subtraction, bad-pixel and flat-field correction. In order to obtain the best possible signal-to-noise ratios and maximum FOV coverages in single maps, we combined all good-quality images obtained in the same night into mosaics.

\subsubsection{Geometric distortion}

In order to avoid systematic alignment errors when mosaicking single images, we corrected the individual frames for the geometric distortion of the CONICA imager. As there is no publicly available description of the instrumental distortion properties of NACO, we extracted the necessary parameters from our data. We modelled the distortion correction using the radially symmetric standard ansatz

\begin{equation}
\hspace{15mm} {\bf r} = {\bf r'}(1-\beta {\bf r'}^2)
\label{eq_distortion}
\end{equation}

\noindent
with

$$
{\bf r} = {\bf x} - {\bf x}_C \hspace{1cm} {\rm and} \hspace{1cm} {\bf r'} = {\bf x'} - {\bf x}_C
$$

\noindent
(e.g. J\"ahne \cite{jaehne2005})\footnote{See also the electronic manual of the public Gemini North Galactic Center Demonstration Science Data Set for an application on GC imaging data.}. Here ${\bf x}$ and ${\bf x'}$ are the true and distorted image coordinates respectively, $\beta$ is a parameter describing the strength of the grid curvature, and ${\bf x}_C\equiv(x_C,y_C)$ is the zero point of the distortion on the detector. Details of the modelling procedures are given in the appendix.

\begin{figure}
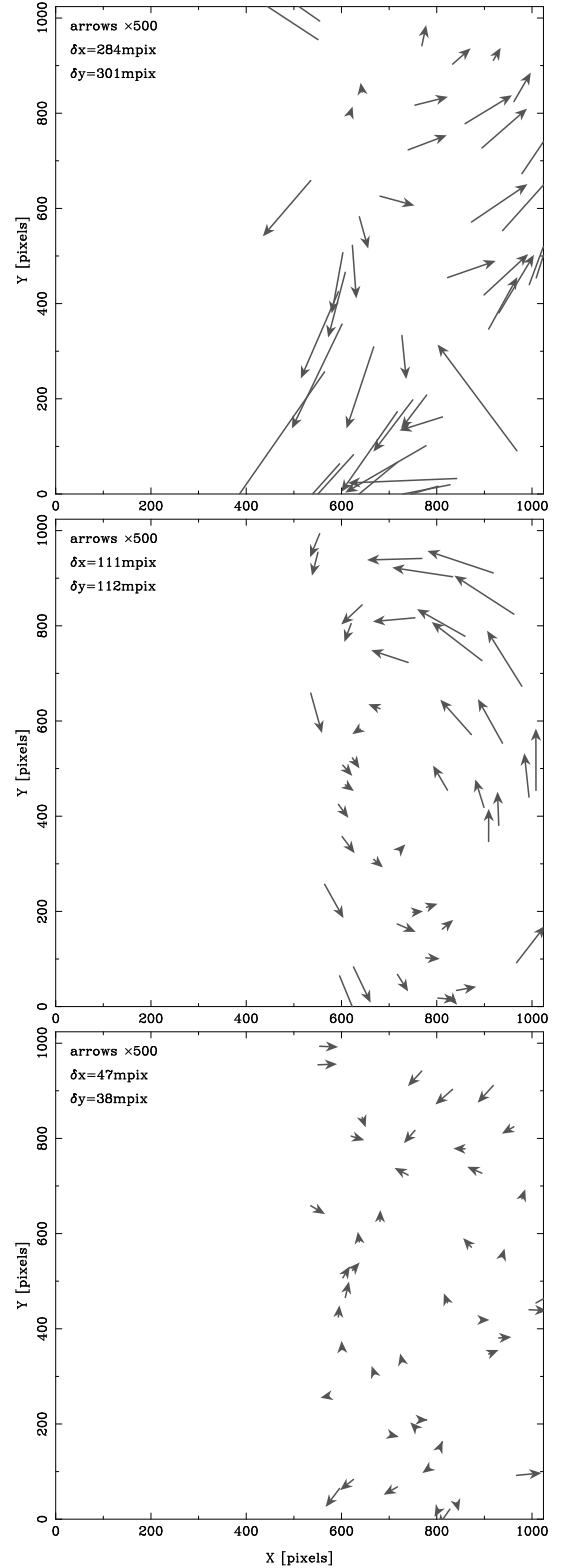
 

\begin{center}

\includegraphics[trim = 0cm 0cm 1cm 0cm, clip, height=7.1cm,angle=-90]{0191f02a.eps}

\includegraphics[trim = 0cm 0cm 1cm 0cm, clip, height=7.1cm,angle=-90]{0191f02b.eps}

\includegraphics[height=7.1cm,angle=-90]{0191f02c.eps}

\end{center}

\caption{Residual image alignment errors before and after distortion correction and image registration, comparing the overlap area (right hand half of the first image of the data set) of two 27~mas/pixel scale K band images obtained in March 2007. Arrows mark absolute values (1~unit = 2~milli-pixels) and directions of residuals, $\delta x$ ($\delta y$) are the rms of residuals in $x$ ($y$). {\it Top panel:} Sub-pixel accurate shift-and-add only. {\it Central panel:} After correcting for geometric distortion, before registration. {\it Bottom panel:} After distortion correction and registration.}

\label{mosaic}

\end{figure}

In case of the \emph{large scale} (27~mas/pixel) images we found
\\

$x_C \simeq 577 \ ... \ 629 \ \ \rm pixels$

$y_C \simeq 775 \ ... \ 823 \ \ \rm pixels$

$\ \beta \ \simeq 2.97 \ ... \ 3.40 \times 10^{-9} \ \ \rm pixels^{-2} ~.$
\\
\\
For the \emph{small scale} (13~mas/pixel) images the parameters were
\\

$x_C \simeq 573 \ ... \ 839 \ \ \rm pixels$

$y_C \simeq 629 \ ... \ 948 \ \ \rm pixels$

$\ \beta \ \simeq 2.06 \ ... \ 13.27 \times 10^{-10} \ \ \rm pixels^{-2} ~.$
\\
\\
For the small scale images the distortion is marginally significant and therefore hard to measure, as evident in the wide range of values found for the curvature $\beta$. Nevertheless, it is possible to state qualitatively that for the large scale images ($\beta\approx3\times 10^{-9}~\rm pixels^{-2}$) the distortion is clearly stronger than for the small scale images ($\beta\approx5\times 10^{-10}~\rm pixels^{-2}$).

\subsubsection{Image registration and mosaicking}

After extracting the distortion parameters, we registered all single frames with respect to a common coordinate grid to ensure alignment to sub-pixel accuracy. After correcting for geometric distortion, no systematic effects of spatial higher-order should remain. In such a case, the image registration can be described by spatial first-order transformations

\begin{eqnarray}
\hspace{10mm} x' = a_0 + a_1 x + a_2 y \\
\hspace{10mm} y' = b_0 + b_1 x + b_2 y
\label{eq_lintrafo}
\end{eqnarray}

\noindent
which cover translations, rotations, scalings, and shears.

The image registration process consists of two main steps. Firstly, we define one master image (usually the first image of a set) as a zero point. We pick a set of bright stars with coordinates $\{{\bf x}_{\rm ref}^0\}$ serving as reference.

Secondly, we find for each image the transformation $T_m$ which transforms the coordinates of the reference stars measured in image $m$ to the coordinates measured in the zero image:

\begin{equation}
\hspace{10mm} T_m: \{{\bf x}_{\rm ref}^m\} \longrightarrow \{{\bf x}_{\rm ref}^0\} ~.
\label{eq_trafo_registration}
\end{equation}

The final mosaic results from computing for each pixel in image $m$ its new position in the mosaic grid (corresponding to image 0) by applying $T_m$. For the transformations we solved the overdetermined sets of equations by means of least-squares fits. We used the routines by Montenbruck \& Pfleger (\cite{montenbruck1989}) implemented in the MPE data analysis software package \emph{DPUSER}\footnote{Developed by Thomas Ott; {\tt http://www.mpe.mpg.de/$\sim$ott/\\dpuser/history.html}}. For each image, the flux values of its pixels were then interpolated to the corresponding mosaic pixels.

Comparing the typical residual alignment errors before and after distortion correction and registration (Fig. \ref{mosaic}) shows clearly the improvement in mosaic quality. While simple shift-and-add leads to inaccuracies as high as some tenths of a pixel, the distortion correction alone provides a strong improvement. When taking into account the grid curvature only, systematic first-order effects (shifts, rotations, and shear) are left. After registering the images, the typical residual errors (pairwise between image overlap areas) are of the order $\approx$0.05 pixels, corresponding to $\approx$1.4~mas (large scale) and $\approx$0.7~mas (small scale), respectively.

\subsection{Spectroscopy}

For obtaining spectroscopic data we used SINFONI, a combination of the integral field spectrometer SPIFFI (Eisenhauer et al. \cite{eisenhauer2003b}, \cite{eisenhauer2003c}) and the adaptive optics system MACAO (Bonnet et al. \cite{bonnet2003}, \cite{bonnet2004}).

SINFONI's data output consists of cubes. The cubes have two spatial axes with dimensions of 64 and 32 pixels respectively and one spectral axis of 2048 pixels length.

Depending on the plate scale, individual cubes covered regions of $0.8\times0.8$'', $3.2\times3.2$'', or $8\times8$''; the latter was used in seeing-limited mode only. The spectra covered either the K band (with a spectral resolution of $R = 4500$) or the band range H+K ($R = 2500$).

SINFONI shows a substantial geometric distortion. This distortion can be parametrized as a 2-dimensional second-order polynomial. We applied the publicly available distortion parameters for correction of our data.

After sky subtraction, bad-pixel-, and flat-field-correction, the wavelength scale was calibrated. This calibration used emission line gas lamps and was finetuned on the atmospheric OH lines. Atmospheric absorption features were removed by dividing by the spectrum of a calibration star.

\section{Astrometry}

\begin{figure*}
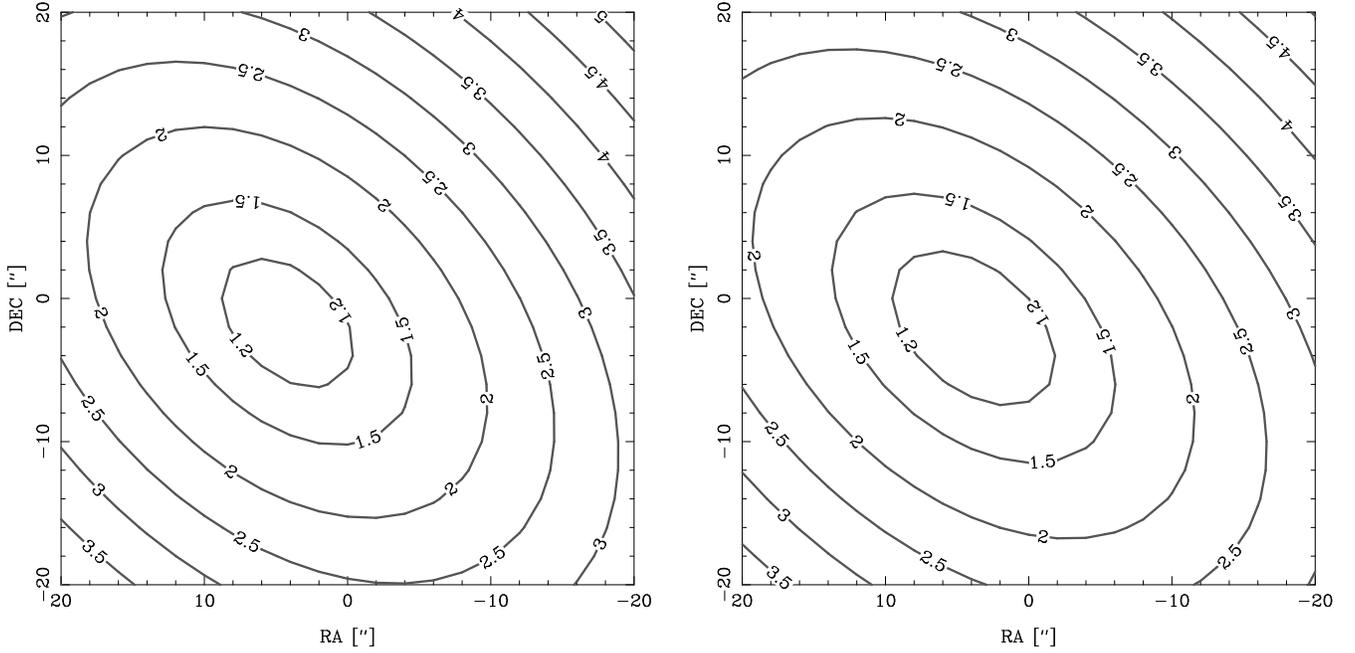
 

\begin{center}

\includegraphics[height=8.6cm, angle=-90]{0191f03a.eps}
\hspace{3mm}
\includegraphics[height=8.6cm,angle=-90]{0191f03b.eps}

\end{center}

\caption{Statistical uncertainty of absolute radio reference frame coordinates as a function of position due to transformation errors, separately for R.A. ({\it left hand panel}) and DEC ({\it right hand panel}). The contour values are given in units of milli-arcseconds. Across the field of view the errors vary in the range 1.2~...~5 mas. The contours mirror the alignment of the reference stars in the sky plane (see Fig. \ref{field}).}

\label{trafo}

\end{figure*}

\subsection{Source selection}

In order to determine positions and proper motions for as many stars as possible, we first constructed a source list from a high-quality large scale K band mosaic with $\approx 40\times 40$'' FOV obtained in May 2005 (epoch 2005.36) . In this image we identified and listed all stars above a given significance threshold using the algorithm {\it FIND} (Stetson \cite{stetson1987}). This algorithm searches an image for positive brightness perturbations and identifies them as stars, if their sharpness and roundness parameters are located within given limits. 

Out of the list of all detected sources we excluded those overlapping (i.e. separated by less than $\approx$2 FWHMs / $\approx$130 mas) with neighbouring stars and thus unusable for precise astrometry. The list of remaining ``good'' stars contains 6037 objects down to magnitudes of K$\approx$18. For all sources the diffraction-limited cores were fit as 2-dimensional elliptical Gaussian brightness distributions.

This procedure of source selection leaves us with a spatial source distribution which is very different from the physical stellar surface density. Due to the restrictions outlined in the previous paragraphs, the sample distribution shows an approximately constant surface density ($\approx$4 sources/arcsec$^2$) across the entire FOV. This does not harm the analysis of kinematic parameters (velocities, velocity dispersions) because there is no bias in velocities. However, in the later stages of this analysis we analyse spatially de-projected parameters such as 3D dispersions. As the outcome of this de-projection crucially depends on the assumed 3D source distribution, we make use of earlier analyses of the 2D and 3D star distributions (Genzel et al. \cite{genzel1996,genzel2000,genzel2003}; Mouawad et al. \cite{mouawad2005}; Sch\"odel et al. \cite{schoedel2007}).

In each individual mosaic $n$ the master list stars were re-identified and their detector positions $\{{\bf X}^n\}$ were fit with 2-dimensional elliptical Gaussian profiles. The formal detector position accuracies were typically (mode of histogram) $\approx$0.025~pixels (per coordinate) in both plate scales, i.e. $\approx$0.68~(0.33)~mas in the large (small) plate scale.

\subsection{Astrometric coordinates}

In order to convert the image positions of the source list stars into absolute astrometric coordinates, we initially use a reference set of 9 SiO maser stars located in the FOV. For these stars absolute positions and motions are known from radio observations (Reid et al. \cite{reid2007}). They are bright sources in both radio and NIR, and are therefore well-suited for cross-calibration.

Although there are no simultaneous observations in the two wavelength regimes, there are radio position measurements from both before and after the epoch of the NIR reference image. The astrometric positions of the SiO maser stars at epoch 2005.36, $\{{\bf x}_{\rm maser}^0\}$, can then be obtained by interpolation (see Reid et al. \cite{reid2007} for details). Using the NIR image detector positions $\{{\bf X}_{\rm maser}^0\}$, we find a transformation

\begin{equation}
\hspace{10mm} T_A: \{{\bf X}_{\rm maser}^0\} \longrightarrow \{{\bf x}_{\rm maser}^0\} ~.
\label{eq_trafo_masers}
\end{equation}

\noindent
By applying $T_A$, we calculate astrometric (reference epoch) positions $\{{\bf x}^0\}$ for all 6037 stars:

\begin{equation}
\hspace{10mm} \{{\bf X}^0\} \longrightarrow T_A\left( \{{\bf X}^0\}\right) = \{{\bf x}^0\} ~.
\label{eq_trafo_masterlist}
\end{equation}

\noindent
The absolute accuracies of these coordinates vary in the range 1.2~...~5~mas depending on position (see Fig. \ref{trafo} for a contour map); details can be found in the appendix.

Unfortunately, in our NIR data no absolute astrometric reference source is available and the SiO maser stars are present only in some of the large scale (27~mas/pix) images. We therefore defined a relative astrometric reference frame tied to an ensemble of $\approx$560 well-behaved (meaning bright and well separated from neighbouring sources) stars with astrometric positions $\{{\bf x}_{\rm ref}^0\} \subset \{{\bf x}^0\}$. For each image $n$ we use the corresponding detector positions $\{{\bf X}_{\rm ref}^n\}$ to compute a linear transformation

\begin{equation}
\hspace{10mm} T_n: \{{\bf X}_{\rm ref}^n\} \longrightarrow \{{\bf x}_{\rm ref}^0\} ~.
\label{eq_trafo_nthimage}
\end{equation}

\noindent
In this step we assume that our reference star ensemble $\{{\bf x}_{\rm ref}^0\}$ is at rest \emph{in average}, i.e. $\langle \{{\bf x}_{\rm ref}^0\} \rangle = \langle \{{\bf x}_{\rm ref}^n\} \rangle$. We note that our procedure is very similar to the local-transformation approach outlined by Anderson et al. (\cite{anderson2006}). The transformation $T_n$ is used to calculate the astrometric positions of \emph{all} stars via

\begin{equation}
\hspace{10mm} \{{\bf X}^n\} \longrightarrow T_n\left( \{{\bf X}^n\}\right) = \{{\bf x}^n\} ~.
\label{eq_trafo_allpos}
\end{equation}

In the \emph{small scale (13 mas/pix) images}, typically only $\approx$100 of the reference stars are present. This means that we only have relatively small subsets $\{{\bf X}_{\rm small}^n\} \subset \{{\bf X}_{\rm ref}^n\}$ at hand for calibration. Therefore we first analyse the large scale (27 mas/pix) images and compute proper motions for all reference stars (see also Sect.~3.3). With the proper motions at hand, we compute for each image $n$ the expected astrometric positions $\{{\bf x}_{\rm small}^n\}$ via linear interpolation. From this we find the (1st order) transformation 

\begin{equation}
\hspace{10mm} T_{n'}: \{{\bf X}_{\rm small}^n\} \longrightarrow \{{\bf x}_{\rm small}^n\}
\label{eq_trafo_smallimage}
\end{equation}

\noindent
which leads to the astrometric positions of \emph{all} stars via

\begin{equation}
\hspace{10mm} \{{\bf X}^n\} \longrightarrow T_{n'}\left( \{{\bf X}^n\}\right) = \{{\bf x}^n\} ~.
\label{eq_trafo_allpossmall}
\end{equation}

\noindent
This procedure ensures that the small scale images are tied to the reference frame of the large scale images.

\subsection{Proper motions}

\begin{figure}
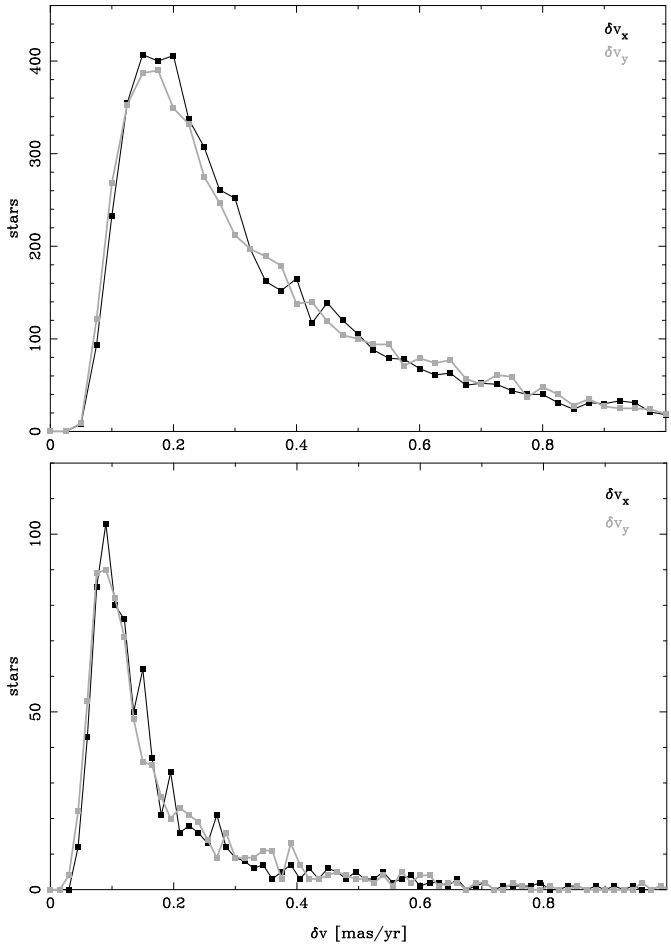
 

\includegraphics[trim = 0cm 0cm 0.9cm 0cm, clip, height=8.8cm,angle=-90]{0191f04a.eps}

\includegraphics[height=8.8cm,angle=-90]{0191f04b.eps}

\caption{Histograms of proper motion errors. {\it Top panel:} Large pixel scale (27~mas/pix). {\it Bottom panel:} Small pixel scale (13~mas/yr). The distributions peak at $\approx$0.18 mas/yr in case of the large pixel scale and at $\approx$0.1 mas/yr in case of the small pixel scale, corresponding to 6.8 km/s and 3.8 km/s respectively.}

\label{velerr}

\end{figure}

We computed stellar proper motions  ${\bf v}_{xy}$ by fitting linear functions to astrometric star positions ${\bf x}$ vs. time $t$:

\begin{equation}
\hspace{10mm} {\bf x}(t) = {\bf v}_{xy} t + {\bf x}(0) ~.
\label{eq_xymotion}
\end{equation}

In order to determine proper errors for the stellar velocities, we applied outlier rejection and error rescaling to the data. The typical (mode of histogram) measurement error of a star position is $\approx$0.9 mas for both image scales. The timeline of observations is five years. The number of epochs is 10 for the large (27~mas/pix) scale and 42 for the small (13~mas/pix) scale imaging data.

In total we were able to extract proper motions for 5548 stars located in the large scale fields; out of these, 755 sources were additionally covered by the small scale images. Typical (mode of histogram) \emph{statistical} proper motion accuracies are $\approx$0.18 mas/yr per coordinate for the large scale data sets and $\approx$0.1 mas/yr per coordinate for the small scale fields. This corresponds to $\approx$6.8 km/s and $\approx$3.8 km/s, respectively. The error distributions are presented in Fig. \ref{velerr}.

We focused on the behaviour of the late-type population of stars which is expected to be dynamically relaxed (Genzel et al. \cite{genzel2003}). Therefore we excluded 103 spectroscopically identified early-type stars which are known to mainly move in disks (Paumard et al. \cite{paumard2006}); this reduced the number of 2D velocity stars from 5548 to 5445. One should however note that probably some more, so far unidentified, early-type stars are still included in our proper motion sample (under investigation by Bartko et al. \emph{in prep.}).

An additional, \emph{systematic} uncertainty is introduced by the relative astrometric reference frame. This frame is based on stars with proper motions known only a posteriori and with respect to the star cluster. Thus a systematic motion of the reference frame is possible. Using the average number of applicable large scale image reference stars, which is 433, and the rms velocity of the reference stars (3.6~mas/yr or 137~km/s), we estimate this systematic uncertainty (standard error) to be 0.17~mas/yr or 6.4~km/s.

\section{Radial velocities}

\begin{figure} 

\includegraphics[height=8.8cm,angle=-90]{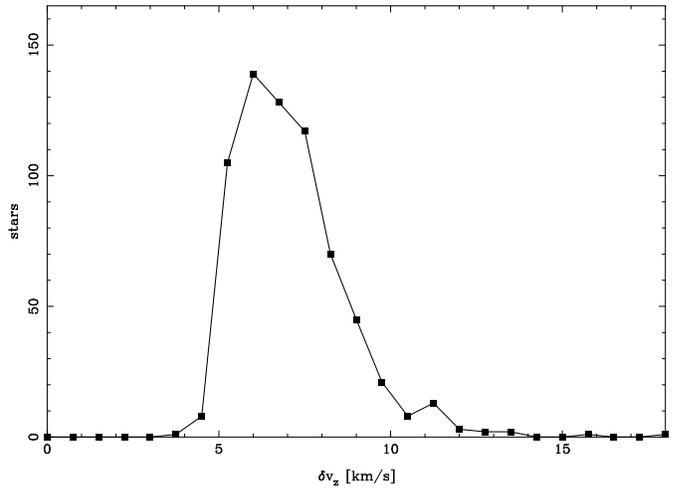}

\caption{Histogram of statistical errors in line-of-sight velocity for 664 stars. The distribution peaks at $\approx$7~km/s.}

\label{vzerr}

\end{figure}

\subsection{Source selection}

We extracted spectra of late-type CO absorption stars from the SINFONI cubes. For each star we selected source and background pixels by hand. Background pixels were selected from pixels surrounding the source pixels. A corrected star spectrum results from subtracting the average of the background pixels from the average of the source pixels. This operation is necessary in order to take into account incomplete sky subtraction, nebular contamination, and flux spillover from neighbouring sources.

Due to the small FOV of SINFONI (8''$\times$8'' at most), our target area (central $\approx$40''$\times$40'') is covered only partially; see Paumard et al. (\cite{paumard2006}) for a detailed overview, especially their Fig.~1. Additionally, the various data sets show different pixel scales, Strehl ratios, and photometric completenesses. This incomplete coverage of the star cluster is a serious limit for 3D de-projection and kinematic phase-space analysis of the data.

\subsection{Velocity fitting}

We extracted stellar radial velocities by correlating the observed spectra with a theoretical template spectrum (e.g. Tonry \& Davis \cite{tonry1979}). The model spectrum of a CO star obtained from the MARCS stellar model-atmosphere and flux library (Gustafsson et al. \cite{gustafsson2003}) served as template. Main model parameters were temperature $T_{\rm eff}=4250$~K, gravitational acceleration $\log g=0$ [cm/s$^2$], micro-turbulence velocity $\vel_{\rm turb}=2$~km/s, and solar metallicities. The model parameters -- especially $T_{\rm eff}$ -- were selected in order to fit the most numerous stars in our sample, which are red clump stars (Maness et al. \cite{maness2007}).

\begin{figure} 

\includegraphics[height=8.8cm,angle=-90]{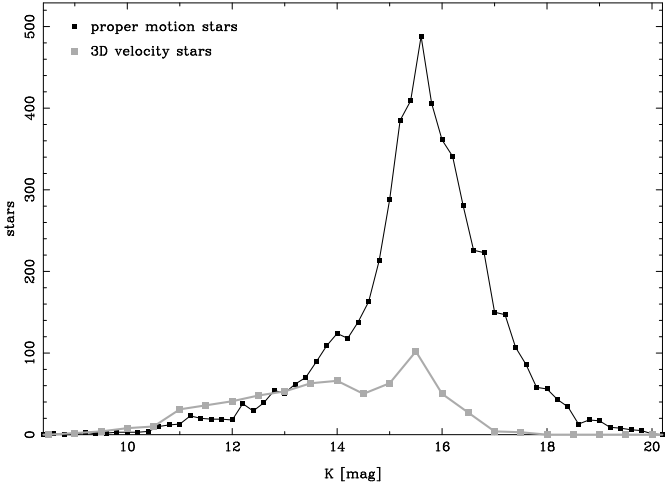}

\caption{K magnitude distributions of all 5445 (non early-type) proper motion stars (black curve) and the 664 3D velocity stars (grey curve). Please note the different binnings. The distributions peak in the range K$\approx$15...16; this is the regime of the Red Clump stars. For K$>$16, the completeness quickly decreases.}

\label{khisto}

\end{figure}

As we were especially interested in the behaviour of evolved late-type stars, we focused our analysis on the CO bandhead lines in the wavelength range $\approx 2.28 ... 2.37~\mu$m. If the maximum correlation was lower than 0.55, the computed velocity was rejected as unreliable. We chose this threshold experimentally after checking the spectra by eye. All velocities were corrected to the local standard of rest using the standard IAU solar motion (see, e.g., Kerr \& Lynden-Bell \cite{kerr1986}, and references therein).

In total we extracted radial velocities for 664 late-type stars. Typical \emph{statistical} velocity accuracies are $\approx$7~km/s; their distribution is presented in Fig.~\ref{vzerr}. Additionally, a \emph{systematic} uncertainty is introduced by the selection of the model spectrum parameters. From a comparison of several model spectra, we estimate this systematic error to be $\approx$5~km/s. Fig.~\ref{khisto} shows the K magnitude distribution of the 5445 proper motion stars and the 664 stars with radial velocities.

\section{Results and discussion}

\subsection{Isotropy and rotation}

With proper motions for 5445 stars and 3D-velocity vectors for 664 stars, we extracted the dynamical properties of the cluster. Our data provide information for projected distances from Sgr~A* up to about 27 arcsec. As a first step we computed velocity dispersions along all coordinate directions.

In order to calculate these parameters we used the method by Hargreaves et al. (\cite{hargreaves1994}). For each coordinate axis $q$, this algorithm computes the average velocity $\langle \vel_q\rangle$ and the velocity dispersion $\sigma_q$ for a given ensemble of stars using the iterative scheme

\begin{equation}
\hspace{15mm} \langle \vel_q \rangle = \frac{\sum_{i}w_{i}\cdot \vel_{q,i}}{\sum_{i} w_{i}}
\label{eq_harg_vavg}
\end{equation}

\begin{equation}
\hspace{15mm} \sigma_q^2 = \frac{\sum_{i} [(\vel_{q,i} - \langle \vel_q \rangle )^2 - \delta_{q,i}^2]\cdot w_{q,i}^2}{\sum_{i} w_{q,i}^2}
\label{eq_harg_sigma}
\end{equation}

\noindent
with $\vel_{q,i}$ being the $q$ component of the velocity of star $i$, $\delta_{q,i}$ being the respective error, and $w_{q,i} = 1 / (\delta_{q,i}^2 + \sigma_q^2)$ being the star's weight. In general, not more than three iterations are necessary to obtain stable results; we usually used five. The respective statistical errors are

\begin{equation}
\hspace{15mm} \delta\langle \vel_q \rangle = \sqrt{\frac{\sigma_q^2 + \langle \delta_{q,i}^2 \rangle}{N}}
\label{eq_harg_deltavavg}
\end{equation}

\begin{equation}
\hspace{15mm} \delta\sigma_q = \frac{\sigma_q^2 + \langle \delta_{q,i}^2 \rangle}{\sigma_q\sqrt{2N}} ~.
\label{eq_harg_deltasigma}
\end{equation}

\noindent
$N$ is the number of stars, $\langle \delta_{q,i}^2 \rangle$ is the mean squared statistical velocity error per star. For the case $\langle \delta_{q,i}^2 \rangle \ll \sigma_q^2$ this simplifies to $\delta\langle \vel_q \rangle \simeq \sigma_q/\sqrt{N}$ and $\delta\sigma_q \simeq \sigma_q/\sqrt{2N}$.

\begin{figure} 

\includegraphics[height=8.8cm,angle=-90]{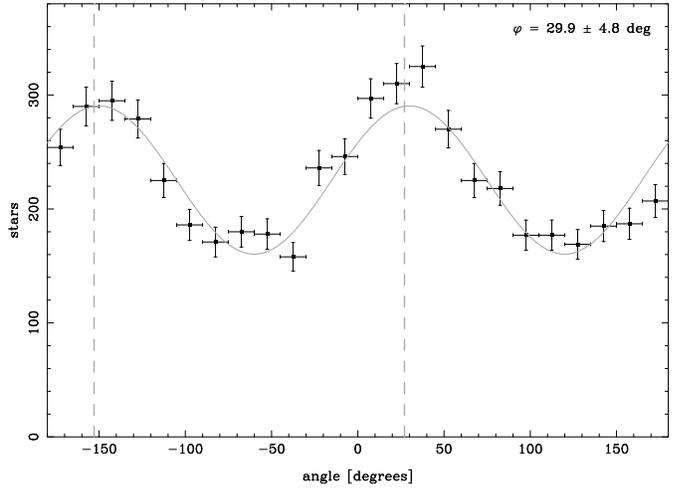}

\caption{Modulations in stellar proper motions. Shown here is the number of stars vs. angle of the proper motion vectors. The angle is defined as atan2($\vel_x$, $\vel_y$) and counted from north to east. Black points with error bars are the data; horizontal error bars mark the full bin widths, vertical error bars are Poisson $\sqrt{N}$ errors. This diagram tests the preferential orientations of proper motion vectors on sky. The vertical dashed grey lines mark the location of the Galactic plane (+27.1$^{\circ}$ and counter-direction). A cosine fit to the data (grey curve) finds a phase $\phi$=+$30\pm5^{\circ}$, in agreement with the orientation of the Galaxy.}

\label{rotation01}

\end{figure}

\begin{figure}
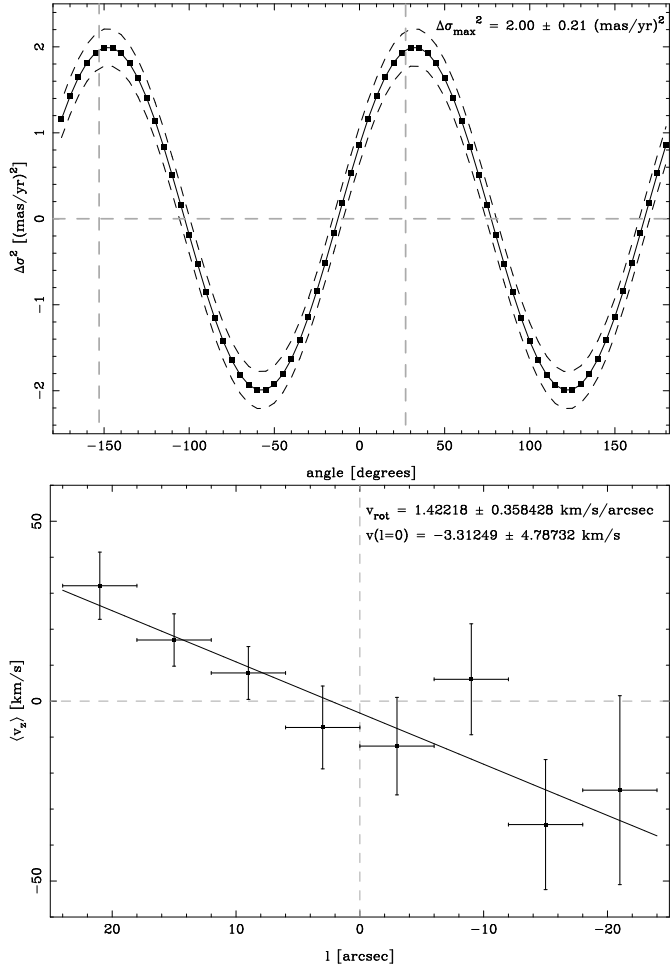
 

\includegraphics[height=8.8cm,angle=-90]{0191f08a.eps}

\includegraphics[height=8.8cm,angle=-90]{0191f08b.eps}

\caption{Signatures of rotation and/or anisotropy in proper motions and radial velocities. {\it Top panel:} Angle on sky (counted from N to E) vs. difference in square dispersions $\Delta\sigma^2 = \sigma_{\parallel}^2 - \sigma_{\perp}^2$. Here $\sigma_{\parallel}$ ($\sigma_{\perp}$) is the velocity dispersion parallel (perpendicular) to a given principal axis which is rotated stepwise. The black dots and the black curve show the observed modulation, dashed black curves mark the 1-$\sigma$ uncertainty range. Vertical grey dashed lines mark the position of the Galactic plane, the horizontal grey dashed line is the zero level of $\Delta\sigma^2$. The modulation has an amplitude of $\Delta\sigma_{\rm max}^2=2.00\pm0.21$~(mas/yr)$^2$. {\it Bottom panel:} Average radial velocities $\langle \vel_z\rangle$ vs. Galactic longitude $l$. Black points are the data, error bars along the $l$ axis mark the full bin sizes, error bars in velocity direction are 1-$\sigma$-errors. Grey dashed lines mark the zero levels of $l$ and $\langle \vel_z\rangle$. As the number of stars is smaller for $l<0$, errors are larger in this range. A linear fit to the data (continuous black line) obtains a rotation velocity of $1.4\pm0.4$ km/s/arcsec.}

\label{rotation02}

\end{figure}

For the two velocity dispersions in R.A. (labelled $x$) and DEC (labelled $y$) using all 5445 proper motion stars we found the values
\\

$\sigma_x = 2.668 \pm 0.027 {\rm \ mas/yr}$

$\sigma_y = 2.824 \pm 0.028 {\rm \ mas/yr}$
\\

\noindent
implying that the dispersions in R.A. and DEC are signicantly different (by about 4$\sigma$).

To check the amount and geometric structure of a possible anisotropy in the proper motion vectors we tested their preferential orientations on sky. For each star we computed the angle $\psi$~=~atan2($\vel_x$,~$\vel_y$) which is counted from north to east\footnote{Where atan2 is the quadrant-preserving arctangent.}. The resulting histogram is shown in Fig. \ref{rotation01}. In case of isotropy the distribution would be flat. The histogram, however, shows a highly significant cosine-like pattern. This pattern is consistent with the signature of a rotating disk seen edge-on, but also with an intrinsic anisotropy in random motions. Fitting this pattern with a cosine profile reveals a phase of +$30\pm 5^{\circ}$, which is in agreement with the plane of the Milky Way located at +27.1$^{\circ}$ (J2000).

\begin{figure} 

\includegraphics[height=8.8cm,angle=-90]{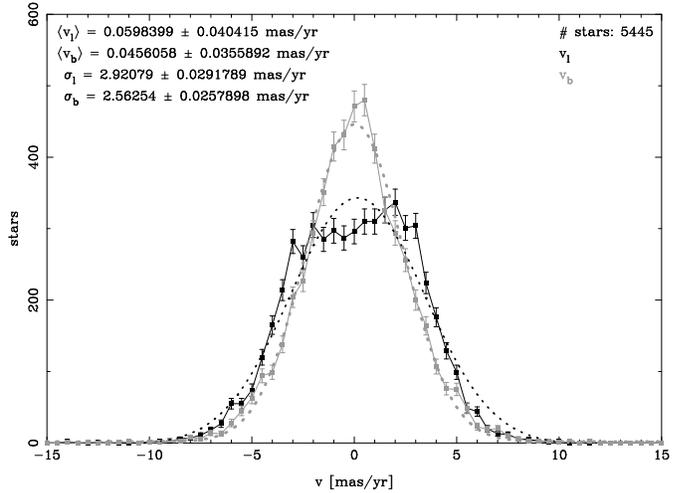}

\caption{Histograms of proper motions in $l$ and $b$ of all proper motion stars. Error bars mark the Poisson errors. Dotted lines represent the respective best-fitting Gaussians. Average velocities $\langle \vel_{l, b}\rangle$ and velocity dispersions $\sigma_{l, b}$ are given in the plot. Whereas the $b$ velocities appear to be normally distributed, the $l$ velocities show a clear rotation pattern.}

\label{dispersions2D}

\end{figure}

The distribution shown in Fig.~\ref{rotation01} shows the presence and orientation of a rotation and/or anisotropy pattern, but not its strength in terms of velocities or velocity dispersions. In order to quantify the modulation of the proper motion distribution, we used the following ansatz: for a given principal coordinate axis we computed the velocity dispersions parallel ($\sigma_{\parallel}$) and perpendicular ($\sigma_{\perp}$) to this axis using all available proper motions. Then we calculated the difference in squares of these two dispersions, $\Delta\sigma^2 = \sigma_{\parallel}^2 - \sigma_{\perp}^2$. By rotating the principal axis stepwise on sky, we obtained $\Delta\sigma^2$ as a function of the angle. The resulting curve, here using a step size of 5$^{\circ}$, is shown in the top panel of Fig. \ref{rotation02}. Since for all data points the same set of proper motions is used, the points are correlated. Using a cosine fit to describe the data, we find an amplitude of $\Delta\sigma_{\rm max}^2=2.00\pm0.21$~(mas/yr)$^2$.

As shown above, the geometry of the cluster kinematics is in good agreement with the orientation of the Galaxy. Thus relative Galactic coordinates $l$, $b$ are a more natural coordinate system than ecliptic coordinates $\alpha$, $\delta$. In the following discussion we will therefore preferentially focus on coordinates and velocities transformed into relative Galactic coordinates, using ($l$,~$b$)$_{\rm Sgr~A*}$~=~(0,~0).

In stellar line-of-sight velocities, rotation in the sense of general Galactic rotation was previously reported (McGinn et al. \cite{mcginn1989}; Genzel et al. \cite{genzel1996} [and references therein]). Therefore we computed for our 664 radial velocity stars (which form a subset of the proper motion stars) the average stellar radial velocities in given $l$ bins. The resulting pattern is shown in the bottom panel of Fig. \ref{rotation02}. Using a linear fit to describe the data points, we find the velocities to be zero within the errors (4.8~km/s) at $l=0$, positive (i.e. receding from the observer) towards positive Galactic longitudes, and negative (i.e. approaching the observer) towards negative $l$, as expected. This does however not imply that the physical (projected) rotation profile actually follows a linear relation; given the limited accuracies of the data, using a more complex rotation model is not justified. Keeping this in mind, we find a rotation velocity of $1.42\pm0.36$ km/s/arcsec. This corresponds to a 4-$\sigma$-detection of the Galactic rotation in radial velocities for $\mid l\mid\le$24~arcsec.

The velocity distributions in $l$ and $b$ for all 5445 proper motion stars are shown in Fig. \ref{dispersions2D} together with the respective best-fitting Gaussian profiles. The velocities in $b$ appear to be normally distributed. In contrast, the histogram of the $l$ velocities shows clear broadening and flattening. The pattern can be approximately described as a convolution of a Gaussian with width $\sigma_b$ and two $\delta$-peaks located at roughly $\pm$2.5~mas/yr. This corresponds to the edge-on view through a system rotating with a fixed rotation velocity of $\approx$2.5~mas/yr. However, this number is an averaged and projected value and affected by the finite FOV; therefore it must not be read as the physical rotation velocity. We will discuss this quantitatively in section 5.3.

\subsection{Distribution of stellar 3D speeds}

Analysing the distribution of the stars' 3D speeds $\vel_{\rm 3D}$ allows one to test wheteher the cluster is dynamically relaxed. In case of the nuclear cluster, the velocity dispersion scales with the projected distance from Sgr~A*, $r$, like $\sigma\propto r^{-0.5}$. As we do not have a detailed dynamical model at hand, we approximate the expected 3D speed distribution as a superposition of \emph{local} (meaning in $r$ bins) Maxwellian distributions. 

In order to test the distribution of the GC star speeds, we analysed 664 stars with known 3D velocities. For each star we computed a bias-corrected\footnote{Velocity squares are limited to values $\ge$0. Therefore the statistical velocity errors systematically shift the results towards higher values.} 3D speed

\begin{equation}
\hspace{10mm} \vel_{\rm 3D} = \sqrt{\vel^2_x + \vel^2_y + \vel^2_z - \delta\vel^2_x - \delta\vel^2_y - \delta \vel^2_z}
\label{eq_3dspeed}
\end{equation}

\noindent
where $\vel_{x,y,z}$ are the velocities (in km/s, assuming $R_0=8$~kpc; see Sect. 1, 5.5) and $\delta\vel_{x,y,z}$ are the respective statistical errors.

In a separate step, we calculated 3D velocity dispersions via

\begin{equation}
\hspace{10mm} \sigma_{\rm 3D}(r) = \sqrt{\sigma^2_x(r) + \sigma^2_y(r) + \sigma^2_z(r)} ~.
\label{eq_3ddispersion}
\end{equation}

\noindent
This $\sigma_{\rm 3D}$ is actually the sum of projected velocity dispersions, meaning it is an approximation for the true 3D dispersion. The average dispersion for all stars is $\langle\sigma_{\rm 3D}\rangle=179\pm5$~km/s. For our analysis, we grouped our stars in $r$ bins of 3 arcseconds width. This bin size ensures that at least 20 stars are located in each bin. For each bin we computed the observed 3D dispersion according to Eq.~\ref{eq_3ddispersion} and from this (and the number of stars in the bin) a Maxwellian profile. The combined profile for all 664 stars is the superposition of all individual Maxwellians.

\begin{figure} 

\includegraphics[height=8.8cm,angle=-90]{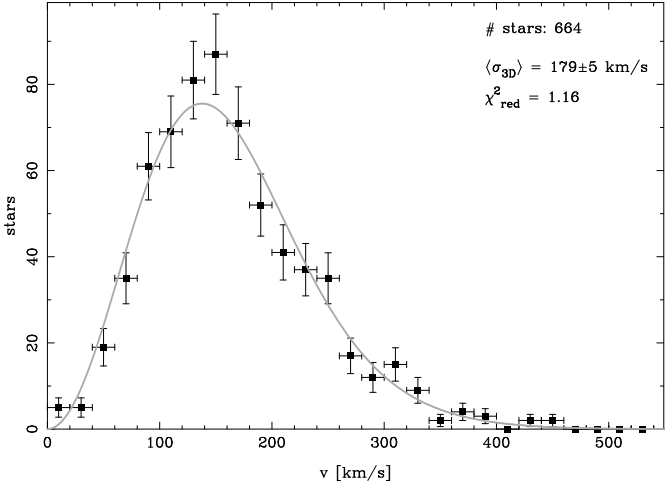}

\caption{Distribution of stellar speeds for all 3D velocity stars. Points with error bars are data; errors in star numbers are Poisson errors, errors in $\vel$ mark the full bin widths. The continuous grey line corresponds to a superposition of Maxwellian distributions for local velocity dispersions $\sigma_{\rm 3D}(r)$; this line is \emph{not} a fit to the data. Observed and model distributions are in good agreement; we find $\chi^2_{\rm red}=1.16$.}

\label{maxwell}

\end{figure}

Observed and theoretical distributions are compared in Fig.~\ref{maxwell}. A reduced-$\chi^2$ test finds $\chi^2_{\rm red}=1.16$, indicating a good agreement. This tells us that our simple model indeed is -- within errors -- a reasonable approximation of the true speed distribution.

Although there is good agreement between the prediction of this model and the data, it is possible that there are few high-velocity stars which are inconsistent with the global distribution (in terms of our Maxwellian approximation). We therefore examined the number of stars with $\vel_{\rm 3D}(r)>2\sigma_{\rm 3D}(r)$. From integrating a Maxwellian, we expect this to be the case for a fraction of 0.0074 of the stars in a given ensemble, i.e. this should be the case only for a very small number of stars. In one case, we find 2 out of the 61 stars located in the respective bin showing such high velocities where 0.45 are expected. Using a Monte Carlo test (with 10,000 realizations) operating on a Maxwellian distribution for 61 stars, we find the probability for two stars having $\vel^2_{\rm 3D}(r)>4\sigma^2_{\rm 3D}(r)$ to be 7.6\% (corresponding to a Gaussian significance of 1.8$\sigma$) -- suggesting this excess is not significant. All stars in our sample are thus compatible with Maxwellian statistics. We therefore conclude that our ensemble of CO stars is consistent we being a uniform, dynamically relaxed system.

This result helps put the high speed of the SiO maser star IRS~9 (cf. Fig.~\ref{field}) recently discussed by Reid et al. (\cite{reid2007}) into context. They found a 3D speed of $\approx$370~km/s for this star located 0.33~pc away from Sgr~A*. They concluded that IRS~9 is too fast to be bound to the mass enclosed within its radial position from Sgr~A*. From our analysis (see Fig.~\ref{maxwell}) one can see that IRS~9 has a high, but not excessive 3D velocity with respect to the global speed distribution. As we find 11 out of 664 (i.e. 1.7\%) stars with speeds above 358~km/s (i.e.$\vel^2_{\rm 3D}>4\langle\sigma^2_{\rm 3D}\rangle$), detecting one out of 15 as in the Reid et al. (\cite{reid2007}) sample does not appear exceptional.

We however find one more star with a very high speed of $\vel_{\rm 3D}=810\pm9$~km/s. This speed exceeds by far the range of the statistical distribution shown in Fig.~\ref{maxwell}. Gillessen et al. (\cite{gillessen2008}) identify this star -- labeled S111 -- as a member of the S-star cluster. Table~\ref{s111} summarizes the properties of S111.

The projected distance $r$ gives a lower limit for the physical 3D distance $R$. From this we compute the highest speed possible for a star which is still bound to the black hole:

\begin{equation}
\hspace{10mm} v_{<\infty} = \sqrt{\frac{2GM_{\bullet}}{r}} ~.
\label{eq_vesc}
\end{equation}

\noindent
Hence we see that S111 might be not bound \emph{to Sgr~A*}. This is in agreement with the findings by Gillessen et al. (\cite{gillessen2008}) who conclude that S111's orbit around Sgr~A* might be hyperbolic. It is however possible that S111 is still bound \emph{to the GC star cluster} if (a) it follows a highly eccentric orbit and (b) we happen to observe it close to its pericentre. In this case, the stellar mass enclosed by the star's orbit can be (together with Sgr~A*) sufficient to bind the star to the cluster. For a more detailed outline of this scenario, see Reid et al. (\cite{reid2007}).

\begin{table} 

\centering

\caption{Properties of the high velocity star S111. Coordinates $x,y$ are given in arcsec, velocities $\vel$ are given in km/s. Errors are statistical.}

\begin{tabular}{l c c}
\hline\hline
parameter & value & error \\
\hline
$x^{\mathrm{a}}$ & -1.127 & 0.001 \\
$y^{\mathrm{a}}$ & -0.936 & 0.001 \\
$r$ & 1.464 & 0.001 \\
$\vel_x$ & -121 & 5 \\
$\vel_y$ & -308 & 5 \\
$\vel_z$ & -739 & 5 \\
$\vel_{\rm 3D}$ & 810 & 9 \\
$\vel_{<\infty}$ & 788 & 40$^{\mathrm{b}}$ \\
\hline
\end{tabular}

\begin{list}{}{}
\item[$^{\mathrm{a}}$] Position fit for reference epoch 2005.36.
\item[$^{\mathrm{b}}$] Error (statistical + systematic) due to uncertainties of $M_{\bullet}$ and $R_0$.
\end{list}

\label{s111}
\end{table}

S111's high velocity might point towards dynamical interactions different from normal two-body relaxation processes in a near-thermal stellar population. In a recent analysis, Perets, Hopman \& Alexander (\cite{perets2007}) point out that the presence of so-called massive perturbers (mainly giant molecular clouds) in the GC region might lead to a substantial number of close encounters between binary stars and Sgr~A*. Such three-body interactions can result in binary disruption with one of the stars being ejected from the GC with a speed up to several thousand km/s (Hills~\cite{hills1988}).

This scenario is of interest especially in view of Galactic hypervelocity stars (HVS) which have speeds higher than the escape speed of the Milky Way (although this is \emph{not} the case for S111 itself) . Right now, 16 of these stars with GC distances in the range $\approx$30...130~kpc are known (Brown et al. \cite{brown2005,brown2007}; Edelmann et al. \cite{edelmann2005}; Hirsch et al. \cite{hirsch2005}; Brown, Geller \& Kenyon \cite{brown2008}). These stars are assumed to have been ejected from the Galactic centre according to the mechanism proposed by Hills~(\cite{hills1988}), although Przybilla et al. (\cite{przybilla2008}) and Bonanos et al. (\cite{bonanos2008}) recently concluded that one of these stars actually originates from the Large Magellanic Cloud. 

In summary, we can conclude the following: (1) the distribution of the stellar 3D speeds confirms the locally relaxed nature of the GC late type cluster. (2) We might have found one high-velocity star unbound to Sgr~A*.

\subsection{Kinematic modelling}

\begin{figure}
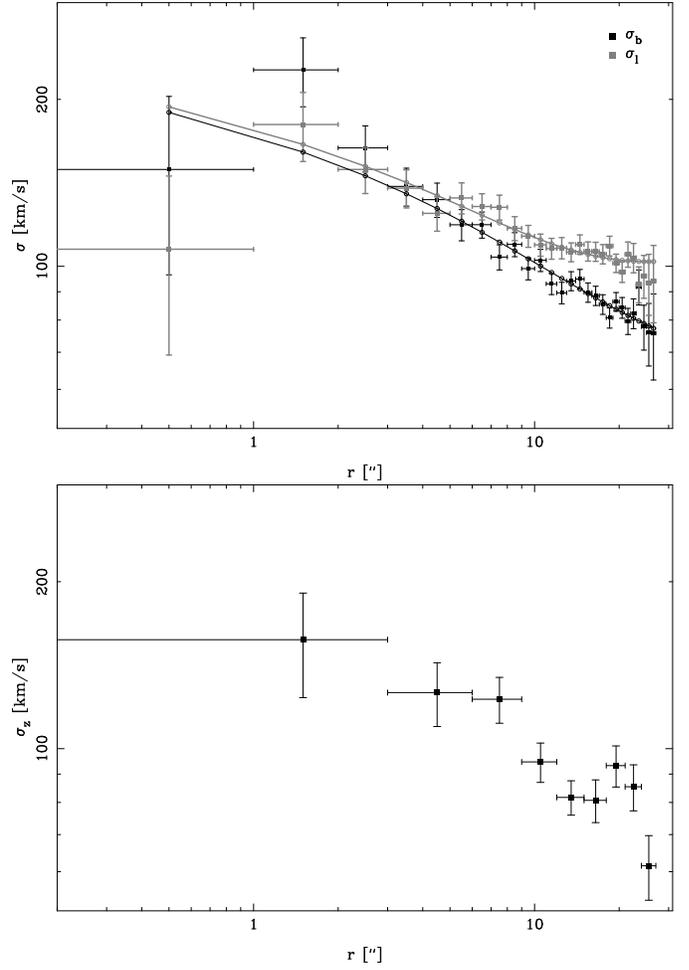
 

\includegraphics[height=8.8cm,angle=-90]{0191f11a.eps}

\includegraphics[height=8.8cm,angle=-90]{0191f11b.eps}

\caption{Velocity dispersion profiles. \emph{Top panel:} Proper motion dispersions. Squares with error bars are observed velocity dispersions $\sigma_{l,b}$ vs. projected distance from Sgr~A* $r$. Continuous lines mark the best-fitting cluster model assuming a spherical, rotating system (please see the text for details). \emph{Bottom panel:} Observed radial velocity dispersions. As the number of 3D velocity stars is about nine times smaller than the number of 2D velocity stars, the profile is sampled sparsely compared to the proper motion dispersion profile.}

\label{dispprofile}

\end{figure}

The large number of proper motions we have at hand allow us to compute a densely sampled velocity dispersion profile. We calculated the velocity dispersions $\sigma_{l,b}$ as functions of projected distance from Sgr~A* $r$ for $r<27$''. The resulting distributions are shown in Fig.~\ref{dispprofile}, (along with the corresponding profile for $\sigma_z$). Within the errors, both profiles ($\sigma_{l,b}$) decrease monotonically with increasing projected distance. With increasing $r$, the two profiles diverge due to rotation and possibly anisotropy.

\begin{figure} 

\includegraphics[height=8.8cm,angle=-90]{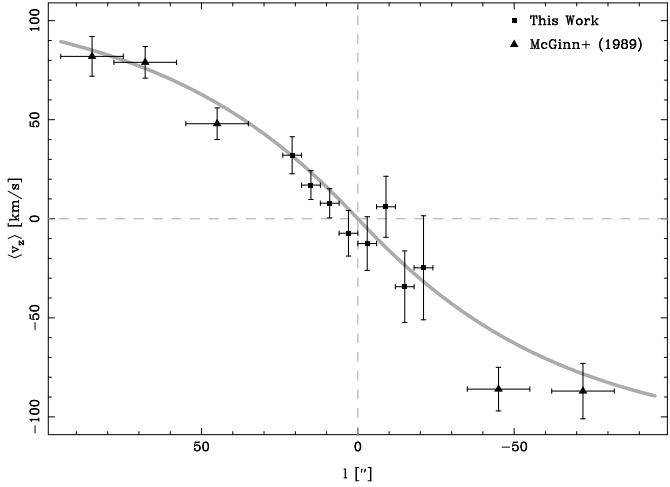}

\caption{Average radial velocity vs. Galactic longitude $l$ as derived from our modelling (continuous grey line). For comparison, the observed values (black data points with error bars) are given. The data are from our work and from McGinn et al. (\cite{mcginn1989}).}

\label{rotationmodel}

\end{figure}

Additional information is provided by the rotation profile $\langle \vel_z \rangle (l)$ obtained from radial velocity data. For $|l|\le$24'' we make use of our SINFONI results (see Fig.~\ref{rotation02} [bottom panel]). We also included data from McGinn et al. (\cite{mcginn1989}) covering the range $|l|=45...85$''; see Fig.~\ref{rotationmodel} for an overview. In the following discussion we neglect the fact that $\langle \vel_z \rangle (l)$ was not measured exactly at $b=0$ but in a strip $|b|\lesssim10$''; we found that this has no significant impact.

In order to quantify the cluster's dynamical properties we simultaneously analysed the proper motion dispersion and rotation profiles. We used an edge-on rotating, spherical stellar system as a model. We fit the model parameters to the data by means of a $\chi^2$ minimization. Coordinates $l, b, z$ are Galactic longitude, Galactic latitude, and l.o.s. axis respectively. Our model contains spherical coordinates $R, \theta , \phi$; these are the 3D distance from Sgr~A*, the zenith angle (with $\theta=0~(\pi)$ being (anti)parallel to the $b$ axis, $\theta=\pi/2$ being parallel to the $l$ axis), and the azimuth angle respectively. The projected 2D distance from Sgr~A* is denoted by $r$.

For our model fit we solve the equations

\begin{equation}
\Sigma (r) \sigma_b^2 (r) = 2\int_r^{\infty}\sigma_b^2(R)\frac{\rho(R) R}{\sqrt{R^2 - r^2}}dR
\label{eq_jeans_sigmab}
\end{equation}

\begin{equation}
\Sigma (r) \sigma_l^2 (r) = 2\int_r^{\infty}\left[\sigma_l^2(R) + \tilde{\vel}_{\phi}(R)^2\left( 1-\frac{r^2}{R^2}\right) \right] \frac{\rho(R) R}{\sqrt{R^2 - r^2}}dR
\label{eq_jeans_sigmal}
\end{equation}

\begin{equation}
\Sigma (r=|l|) \vel_z(|l|) = 2\int_{r=|l|}^{\infty}\vel_{\phi}(R,\theta=\pi/2)\frac{r}{R}\frac{\rho(R) R}{\sqrt{R^2 - r^2}}dR
\label{eq_jeans_vzl}
\end{equation}

\noindent
with

\begin{equation}
\Sigma (r) = 2\int_r^{\infty} \rho(R)\frac{R}{\sqrt{R^2 - r^2}}dR
\label{eq_jeans_density} ~.
\end{equation}

\noindent
$\Sigma (r)$ is the cluster's stellar surface mass density in the sky plane (not the one of the sample population; see the discussion in Section 3.1), $\rho(R)$ is the 3D stellar volume mass density, $\sigma_{l,b}(R)$ is the intrinsic velocity dispersion, and $\vel_{\phi}(R,\theta)$ is the rotation speed (e.g. Wilson \cite{wilson1975}; Binney \& Tremaine \cite{binney1987}; Gerhard \cite{gerhard1994}; Genzel et al. \cite{genzel1996,genzel2000}).

In the integrations we use the parametrized functions

\begin{equation}
\hspace{10mm} \vel_{\phi}(R,\theta) = \vel_r\frac{R\sin\theta}{\sqrt{R^2\sin^2\theta + \eta ^2}}
\label{eq_jeans_rotation01}
\end{equation}

\begin{equation}
\hspace{10mm} \tilde{\vel}_{\phi}(R) = \frac{1}{\pi}\int_0^{\pi} \vel (R,\theta) d\theta
\label{eq_jeans_rotation02}
\end{equation}

\begin{equation}
\hspace{10mm} \rho(R) = \frac{\rho_0}{1+(R/R_b)^2}
\label{eq_jeans_nr}
\end{equation}

\begin{equation}
\hspace{10mm} \sigma_b^2(R) = \sigma^2_0 (R/R_z)^{2\alpha} + \sigma^2_{\infty,b}
\label{eq_jeans_sigr_b}
\end{equation}

\begin{equation}
\hspace{10mm} \sigma_l^2(R) = \sigma^2_0 (R/R_z)^{2\alpha} + \sigma^2_{\infty,l}
\label{eq_jeans_sigr_l} ~.
\end{equation}

\noindent
In this parametrization, $\rho(R)$ is given as a powerlaw sphere profile with break radius $R_b$; $\sigma_{l,b}(R)$ is given as a superposition of a powerlaw dispersion profile and a -- possibly anisotropic -- constant floor dispersion at infinite distances $\sigma_{\infty,l,b}$. The rotation velocity is described by a profile rising from 0 to an asymptotic speed $\vel_r$ with a characteristic scale $\eta$.

We selected the parametrization of the density profile according to the results from observations of stellar number counts  and K-band surface brightness distributions\footnote{In this discussion, we assume a constant mass-to-light ratio.} (Launhardt et al. \cite{launhardt2002}; Genzel et al. \cite{genzel2003}; Mouawad et al. \cite{mouawad2005}; Sch\"odel et al. \cite{schoedel2007}). However, the characteristic scale $\eta$ is not well constrained by the observations; reported values range from $\approx$6'' (Launhardt et al. \cite{launhardt2002}) to $\approx$10'' (Genzel et al. \cite{genzel2003}). Therefore we introduce $\eta$ as a fit parameter into our model because it is not a priori clear that the $\eta$ which fits best the kinematics is the same as the density profile break radius.

Since for our purposes the normalization factors $\rho_0$, $R_z$ (Eq.~\ref{eq_jeans_nr}, \ref{eq_jeans_sigr_b}, \ref{eq_jeans_sigr_l}) are arbitrary, we fixed them to 1. It is well-known that the cluster is dominated by the point mass Sgr~A*; we therefore selected a Keplerian profile for the velocity dispersion, i.e. $\alpha=-0.5$.

In summary, we fit for the 6 fit parameters $\{ \sigma_0, \sigma_{\infty,b}, \sigma_{\infty,l}, R_b, \vel_r, \eta \}$. We used the \emph{Mathematica FindMinimum}\footnote{Wolfram Research, Inc., Champaign, IL, USA} routine to fit the model to the observed dispersion profiles $\sigma_{l,b}(r)$. From the best-fitting model we found
\\

$\sigma_0 = 359 \pm 30~\rm km/s$

$\sigma_{\infty,b} = 54.2 \pm 8.0~\rm km/s$

$\sigma_{\infty,l} = 55.4 \pm 8.8~\rm km/s$

$R_b = 8.9 \pm 3.5~\rm arcsec$

$\vel_r = 189 \pm 38~\rm km/s$

$\eta = 109 \pm 44~\rm arcsec ~.$
\\
\\
Parameter errors are statistical (68\% confidence level, $\Delta\chi^2=7.17$). The fact that some of these errors are relatively large mirrors correlations between parameters. We find $\chi^2_{\rm red}\equiv\chi^2/F=0.79$; $F=61$ is the number of degrees of freedom. This shows a good agreement between model and observational data. The resulting model profiles are presented in Figs.~\ref{dispprofile},~\ref{rotationmodel} together with the corresponding data.

\begin{figure} 

\includegraphics[height=8.8cm,angle=-90]{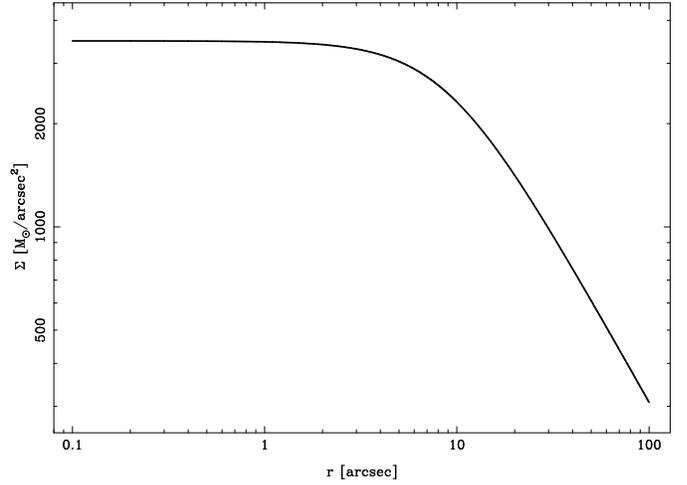}

\caption{Stellar surface mass density $\Sigma$ vs. projected distance $r$ from Sgr~A* for the model cluster.}

\label{modeldensity}

\end{figure}

We can now draw several conclusions. The main assumptions our model is based on, especially stellar density profile and velocity dispersion profiles, actually allow for a reasonable fit ($\chi^2_{\rm red}\lesssim1$). We can thus be confident that our model description is sufficiently complete. Our result $R_b\approx9$'' is in good agreement with the observations of the density distribution. Fig.~\ref{modeldensity} shows the stellar surface density as found by our modelling. The different scalings (mass densities vs. observed number densities) aside, this profile is in good agreement with the earlier observations discussed above.

As $\sigma_{\infty,l}=\sigma_{\infty,b}$ within the errors, the nuclear cluster can be described as an isotropic system.

Until now, our discussion was focused on the kinematic description of the nuclear cluster. In the following, we will draw dynamical conclusions.

Since we have analytic parametrizations for the stellar density profile (Eq.~\ref{eq_jeans_nr}), dispersion profile (Eq.~\ref{eq_jeans_sigr_b}, \ref{eq_jeans_sigr_l}), and rotation profile (Eq.~\ref{eq_jeans_rotation01}, \ref{eq_jeans_rotation02}) at hand, we can compute the mass profile of the cluster. For a spherical, isotropic, rotating system, the mass distribution is given by the Jeans equation via

\begin{equation}
\hspace{7mm} G M(R) = -R \sigma^2 \left( \frac{d\log\rho}{d\log R} + \frac{d\log\sigma^2}{d\log R} - \frac{\vel_{\phi}^2}{\sigma^2} \right)_R
\label{eq_jeans_encmass}
\end{equation}

\noindent
(e.g. Binney \& Tremaine \cite{binney1987}). This relation allows us to compute the enclosed mass as a function of $R$. A non-zero value at $R=0$ should correspond to a central point mass, in our case the mass of Sgr~A*. 

\begin{figure} 

\includegraphics[height=8.8cm,angle=-90]{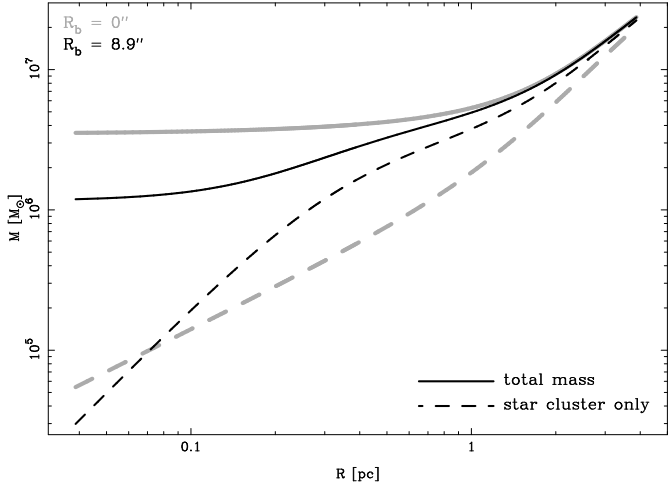}

\caption{Enclosed mass vs. distance from Sgr~A* calculated via the Jeans equation for $1''<R<100''$. The \emph{continuous black} curve is the mass profile using our best-fitting model density profile with break radius $R_b=8.9''$. The \emph{continuous grey} line shows the profile obtained when neglecting the flattening of the density profile towards small $R$, i.e. assuming $R_b=0$. The corresponding \emph{dashed} curves are the stellar mass profiles obtained by subtracting the central point masses (i.e. the values found at $R=0$). The scale is 1~pc~$\equiv$~25.8''.}

\label{massprofile}

\end{figure}

In Fig.~\ref{massprofile} we show the results found when inserting our best-fitting model parameters (black curves). Contrary to the known value $M_{\bullet}\simeq4\times10^{6}M_{\odot}$ (Sch\"odel et al. \cite{schoedel2002,schoedel2003}; Ghez et al. \cite{ghez2003, ghez2005, ghez2008}; Eisenhauer et al. \cite{eisenhauer2005}; Gillessen et al. \cite{gillessen2008}), the total enclosed mass drops down to $M(R\rightarrow0)\simeq1.2\times10^{6}M_{\odot}$. The formal statistical $1\sigma$ uncertainty of this value is $\approx$15\%, meaning the discrepancy is systematic.

This inconsistency in the behaviour of Jeans equation mass profiles in star cluster cores was already noted by Kormendy \& Richstone (\cite{kormendy1995}). They conclude that a good sensitivity to central point masses is given only in the case of steep ($d\log\rho/d\log R\lesssim -2$) density profiles. If the central core of a stellar system is resolved (leading to $d\log\rho/d\log R\approx 0$ for $R\rightarrow 0$), the enclosed mass drops systematically compared to measurements using the large-scale, ``intrinsic'' slope of the density profile. This behaviour is obvious from Eq.~\ref{eq_jeans_encmass}. We therefore might interpret our findings as due to a ``resolution effect''. This is especially interesting in view of the fact that the central few arcseconds of the nuclear cluster are dominated by a different stellar population. Whereas we analyse the properties of CO late-type stars, towards small $R$ an increasing fraction of the stars belongs to a population of young early-type stars. Therefore the intrinsic slope of the density profile in the central part of the cluster is not obvious at all.

In order to check the influence of this ``resolution effect'', we repeated our calculations neglecting the flattening of the density profile towards the cluster centre, i.e. assuming $R_b=0$. The resulting mass profiles are shown in Fig.~\ref{massprofile} (grey curves). Here we indeed find $M(R\rightarrow0)\simeq3.6\times10^{6}M_{\odot}$, in agreement with the mass of Sgr~A* (within the errors). For $R>1$pc, the discrepancy between the profiles vanishes. However, we do not conclude from this that we have found the true mass of the central SMBH by applying the proper assumption. Instead, the range of results should be read as the systematic uncertainty intrinsic to this type of mass estimates. This means that the Jeans profile ansatz allows us to derive the correct order of magnitude (a few $10^{6}M_{\odot}$ in our case) of the central point mass, but with a factor $\approx$2 systematic uncertainty.

In any case ($R_b=8.9''$ or $R_b=0$), we can derive the enclosed \emph{stellar} masses $M_*(R)$ by subtracting the corresponding central point masses. The resulting profiles are given in Fig.~\ref{massprofile} (black and grey dashed curves). Within the systematic uncertainties, the profiles begin at $M_*(R=1'')\simeq4\times10^4 M_{\odot}$ and rise to $M_*(R=100'')\simeq2.3\times10^7 M_{\odot}$. Although this is just consistent with earlier studies (e.g. Lindqvist, Habing \& Winnberg \cite{lindqvist1992}; Ghez et al. \cite{ghez1998}; Genzel et al. \cite{genzel1996,genzel2000} [and references therein]), our masses are systematically higher by factors $\approx$1.5. This is due to the fact that we find a faster rotation of the cluster than the aforementioned works. For the same reason, we find a somewhat smaller sphere of influence (meaning the radius where $M_*(R)=M_{\bullet}$; Alexander (\cite{alexander2005})) for Sgr~A* which is $\approx$1.5~pc (instead of $\approx$3~pc; Alexander (\cite{alexander2005})).

In order to calibrate the $\Sigma$-axis of Fig.~\ref{modeldensity}, we adopted the following approach. We use Eq.~\ref{eq_jeans_encmass} to compute the enclosed mass $M(R)$ at a reference distance $R$ which is large enough to avoid the range where the mass profile shows substantial uncertainties but still within the range covered by data (see Fig.~\ref{rotationmodel}). We picked $R=47$''. For this $R$, the expressions given in the $(\cdots)$ part of Eq.~\ref{eq_jeans_encmass} are $\approx-1.9$, $\approx-0.5$, and $\approx1.0$, respectively. From this we find $M(R=47'')\simeq8.3\times10^6 M_{\odot}$. Subtracting the contribution by Sgr~A* ($M_{\bullet}\simeq4\times10^{6}M_{\odot}$) leaves us with an enclosed stellar mass $M_*(R=47'')\simeq4.3\times10^6 M_{\odot}$. With this value, we obtain $\rho_0\simeq2.1\times10^6 M_{\odot}/{\rm pc}^3$ (see Eq.~\ref{eq_jeans_nr}).

\subsection{Phase-space distributions}

\begin{figure*}
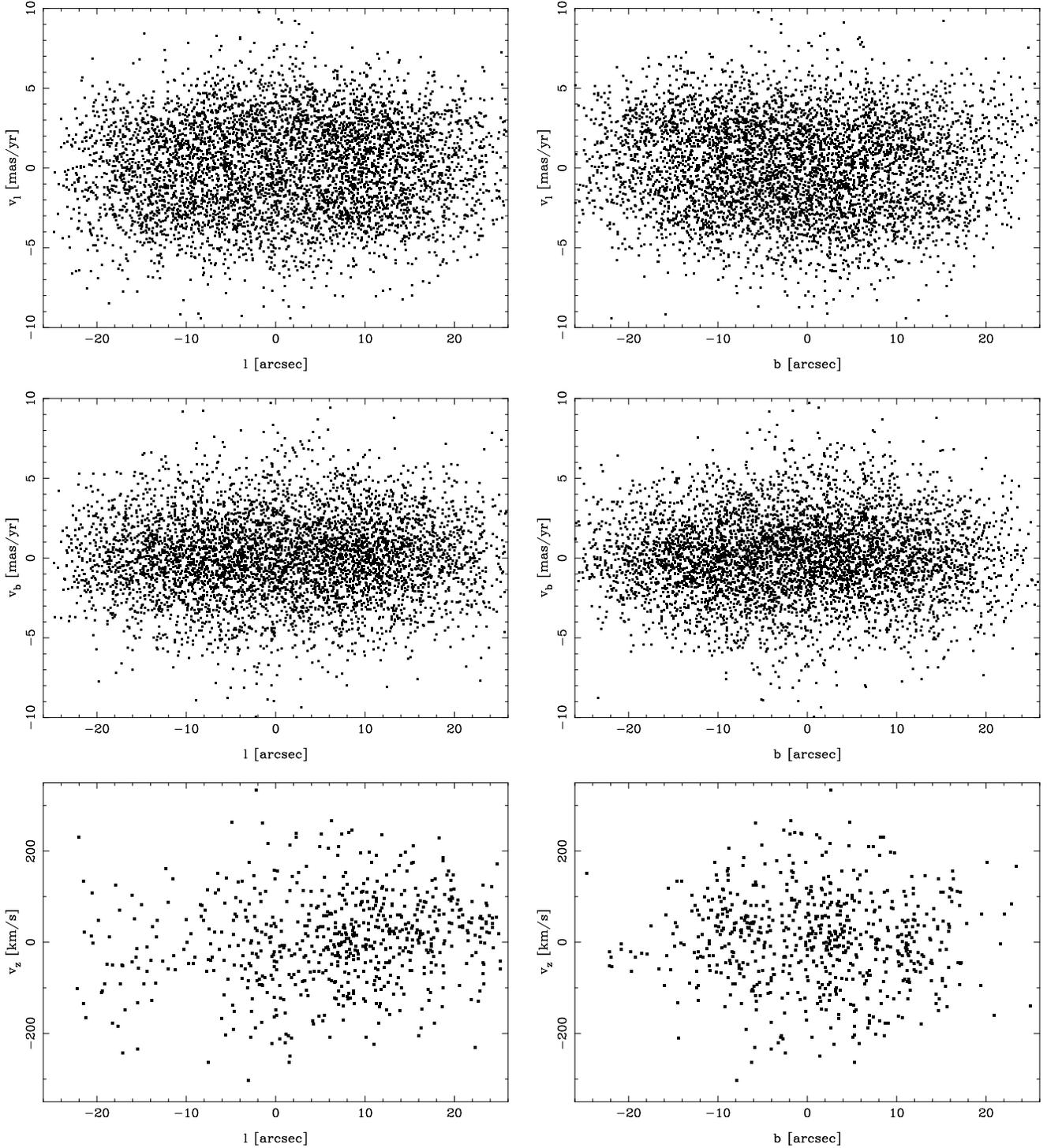
 

\centering

\includegraphics[height=8.5cm,angle=-90]{0191f15a.eps}
\hspace{3mm}
\includegraphics[height=8.5cm,angle=-90]{0191f15b.eps}

\vspace{3mm}

\includegraphics[height=8.5cm,angle=-90]{0191f15c.eps}
\hspace{3mm}
\includegraphics[height=8.5cm,angle=-90]{0191f15d.eps}

\vspace{3mm}

\includegraphics[height=8.5cm,angle=-90]{0191f15e.eps}
\hspace{3mm}
\includegraphics[height=8.5cm,angle=-90]{0191f15f.eps}

\caption{Velocity-coordinate phase-space maps. Velocities in the sky plane $\vel_{l,b}$ include all 5445 proper motion stars; line-of-sight velocities $\vel_z$ are given for 664 stars. The scale is 1~mas/yr~$\equiv$~37.9~km/s. Compared to the $\vel_b$, the $\vel_l$ distributions are broadened (in velocity) due to the global rotation. The $\vel_z$ are biased towards positive $l$ as SINFONI spectra were collected mainly north of Sgr~A*. These global properties aside, the diagrams show no obvious patterns or sub-structures.}

\label{phasespaces}

\end{figure*}

%%%%%%%%%%%%%%%%%%%%%%%%%%%%%%%%%%%%%%%%%%%%%%%%%%%%%

\begin{figure*}
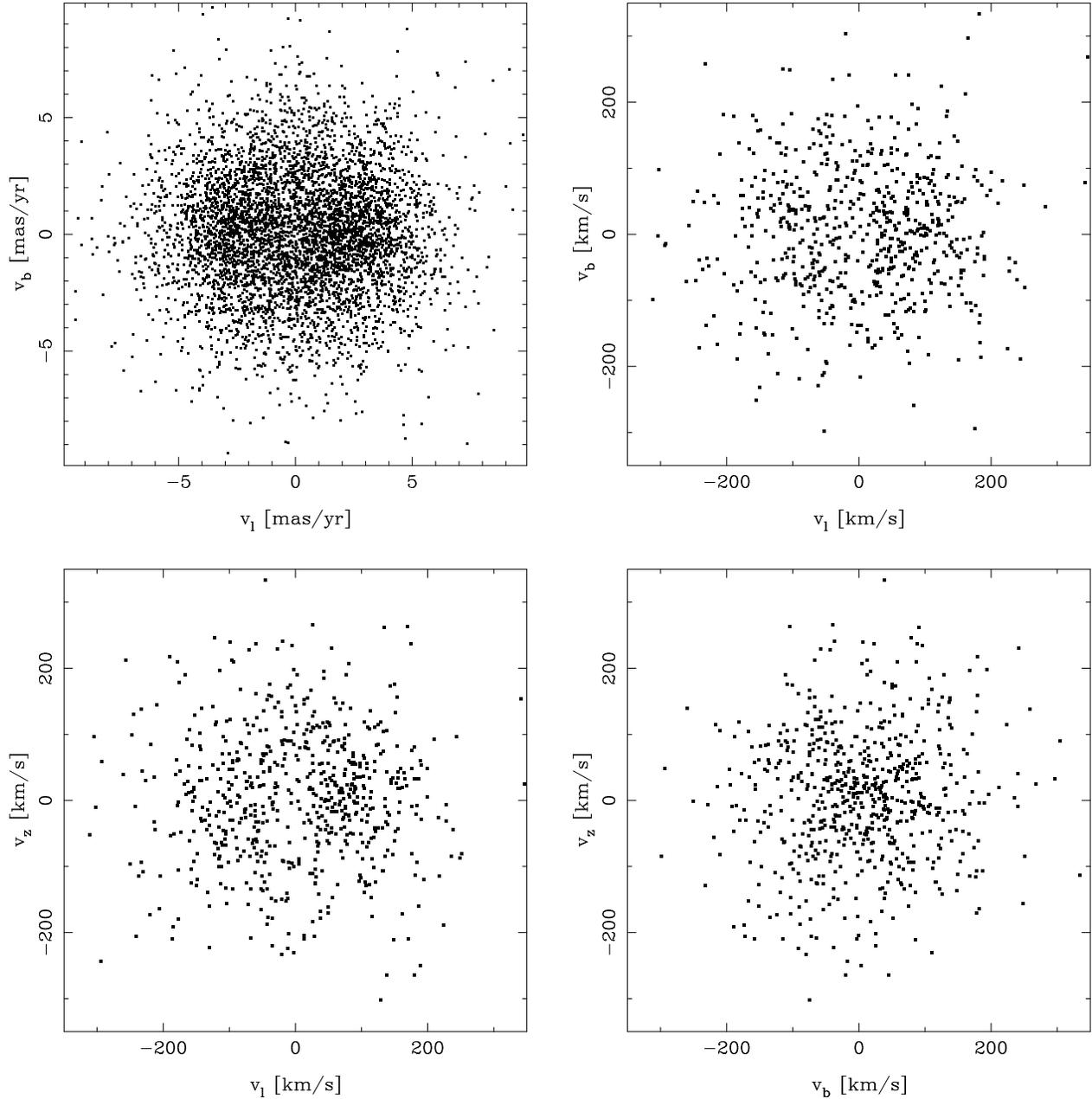
 

\centering

\includegraphics[height=8cm,angle=-90]{0191f16a.eps}
\hspace{5mm}
\includegraphics[height=8cm,angle=-90]{0191f16b.eps}

\vspace{5mm}

\includegraphics[height=8cm,angle=-90]{0191f16c.eps}
\hspace{5mm}
\includegraphics[height=8cm,angle=-90]{0191f16d.eps}

\caption{Velocity-velocity phase-space maps. The distribution of $\vel_l$ vs. $\vel_b$ is given twice, once for all 5445 proper motion stars (\emph{top left}) in mas/yr, once for all 664 3D motion stars (\emph{top right}) in km/s. The scale is 1~mas/yr~$\equiv$~37.9~km/s. In analogy to Fig.~\ref{phasespaces}, the global rotation shows up as a broadening of the $\vel_{l,z}$ distributions with respect to $\vel_b$ (see also Figs.~\ref{dispersions2D}, \ref{dispersions3D}). These global properties aside, there is no obvious substructure or grouping.}

\label{phasespaces2}

\end{figure*}

%%%%%%%%%%%%%%%%%%%%%%%%%%%%%%%%%%%%%%%%%%%%%%%%%%%%%

\begin{figure*}
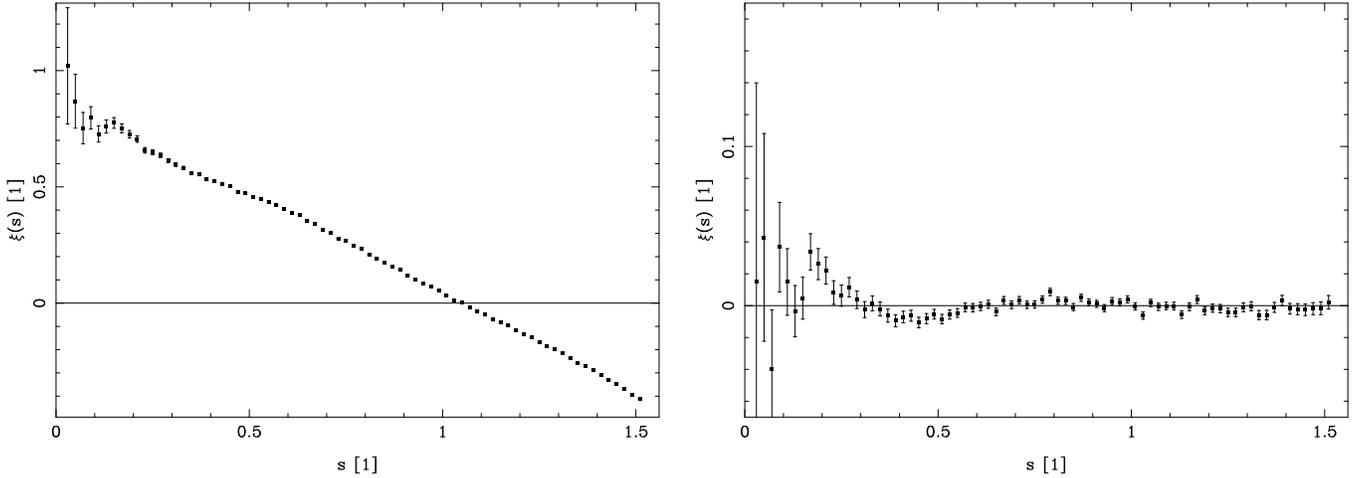
 

\centering

\includegraphics[height=8.7cm,angle=-90]{0191f17a.eps}
\hspace{3mm}
\includegraphics[height=8.7cm,angle=-90]{0191f17b.eps}

\caption{4D two-point correlation functions for all proper motion stars. The normalized distance $s$ is defined via $s^2=(\Delta l/x_0)^2 + (\Delta b/x_0)^2 + (\Delta \vel_l/v_0)^2 + (\Delta \vel_b/v_0)^2$; $x_0$, $\vel_0$ are constants. Please note the different $\xi(s)$ axis scales. \emph{Left hand panel}: Measured phase-space distances compared to a uniform random distribution. The profile mirrors the obvious clustering of phase-space points around $(l,b,\vel_l,\vel_b)=(0,0,0,0)$ shown in Figs.~\ref{phasespaces},\ref{phasespaces2}. \emph{Right hand panel}: Measured phase-space distances compared to a Monte-Carlo model of the GC cluster. Deviations from the uniform rotator model cluster do not exceed $\approx$0.04.}

\label{tpcf}

\end{figure*}

As shown in sections 5.1--5.3, the observed part of the GC star cluster can approximately be described as a spherical rotator with normally distributed random stellar velocities. It is, however, not clear if the observed stars kinematically indeed form a single system. A well-known example for kinematic segregation in the GC is the dichotomy between isotropically distributed old late-type stars and young early-type stars arranged in disks (Genzel et al. \cite{genzel2003}; Paumard et al. \cite{paumard2006}).

A common way to characterize a stellar system is the use of phase-space maps (e.g. Ibata et al. \cite{ibata2001}; Yanny et al. \cite{yanny2003}; Martinez-Delgado et al. \cite{martinez2004}; Seabroke \& Gilmore \cite{seabroke2007}). Applying this method to our data, we construct a variety of phase-space diagrams. The resulting distributions are presented in Figs.~\ref{phasespaces} and \ref{phasespaces2}. Qualitatively, there appears to be no substructure or grouping. In the $\vel_z$-$l$-diagram the data are biased towards positive $l$. This is an observational artefact as SINFONI spectra were preferentially obtained north of Sgr~A*\footnote{Because SINFONI's AO guide star is located $\approx$17'' north of Sgr~A*.}, roughly corresponding to $l>0$.

The phase-space maps also mirror the influence of global rotation and anisotropy discussed in the previous subsections. These effects show up as a broadening of the $\vel_l$-vs-coordinate distributions along the velocity axes compared to the respective distributions for $\vel_b$.
However, one can hardly recognize the rotation pattern from the $\vel_z$-$l$ plot as the random scatter of the data points (i.e. the dispersion) is much larger than the modulation in the velocity average (cf. Fig.~\ref{rotation02} [bottom panel]).

In order to quantify the presence (or absence) of phase-space substructure, we made use of the two-point correlation function (TPCF)

\begin{equation}
\hspace{20mm} \xi(s) = \frac{n_R}{n_D} \frac{DD(s)}{DR(s)} - 1
\label{eq_tpcf}
\end{equation}

\noindent
(Davis \& Peebles \cite{davis1983}). For a given distance $s$, $DD(s)$ is the number of pairwise distances between observed stars (also referred to as the data--data distances) located in the corresponding distance bin. $DR(s)$ is the number of pairwise distances between the observed stars and the members of a comparison ensemble, usually a random, uniform one (thus $DR(s)$ is also referred to as the data--random distances). $n_D\ (n_R)$ is the total number of data--data (data--random) distances. By definition, $\xi(s)$ is located in the range $[-1;+\infty]$; $\xi(s)=0$ corresponds to full agreement between observed ensemble and comparison ensemble.

We computed the TPCF for all 5445 proper motion stars. Since, in phase-space, the distance parameter $s$ mixes positions and velocities, we used a normalized 4D distance 

\begin{equation}
\hspace{7mm} s=\sqrt{(\Delta l/x_0)^2 + (\Delta b/x_0)^2 + (\Delta \vel_l/\vel_0)^2 + (\Delta \vel_b/\vel_0)^2} ~.
\label{eq_4ddistance}
\end{equation}

\noindent
Here $x_0, \vel_0$ are (a priori arbitrary) constant distances and velocities; $\Delta$ denotes the difference in the given coordinate. In order to match the phase-space dimensions of the cluster (cf. Figs.~\ref{phasespaces},~\ref{phasespaces2}), we chose $x_0=1~{\rm pc}=25.8''$, $\vel_0=400~{\rm km/s}=10.55$~mas/yr (with $R_0=8$~kpc). Thus $s=1$ corresponds to the half side length of a ``phase space unit cell''.

As a first step, we computed the TPCF using a random, uniformly distributed comparison ensemble. The resulting distribution is shown in Fig.~\ref{tpcf} (left hand panel). It mirrors the obvious clustering of phase-space points around $(l,b,\vel_l,\vel_b)=(0,0,0,0)$ shown in Figs.~\ref{phasespaces},~\ref{phasespaces2}. The point where the profile crosses the $\xi(s)=0$ line can be identified as a characteristic phase-space radius of the cluster; this radius is $s_c\simeq1.05$.

As a second step, we computed the TPCF using a Monte Carlo model of the cluster. This model is a random realization of the kinematic model solution described in the previous section. We extracted model data by applying an ``on-sky'' selection mask simulating the actual imaging observations and the selection of stars (FOV, minimum star-star distances). These values were inserted into the TPCF calculation.

The right hand panel of Fig.~\ref{tpcf} shows the resulting $\xi(s)$ profile. In general, data and model are in very good agreement, deviations are typically of the order $<$1\% (from $\xi(s)+1=1$). The largest deviations from data-model equality, about 4\%, occur at $s\approx0.15$. Such a signal at small $s$ corresponds to a slight excess of stars with small pairwise phase-space distances compared to the model distribution. This would mean that either (1) the model distribution slightly underestimates the number of stars with small $s$ or (2) the cluster contains a small excess (with respect to a random sample) population of stars moving coherently. However, in either case it is safe to conclude that our -- quite simple -- kinematic model indeed reproduces the observed projected phase-space distribution of the GC cluster.

We did not include stellar radial velocities in the TPCF analysis. The spectra were extracted from 24 separate SINFONI data sets with different FOVs, pointings, pixel scales, integration times, spectral ranges, PSFs, and limiting magnitudes (see Sect.~4.1). This prevented a consistent reconstruction of the observation/selection mask, thus excluding the reliable extraction of a model cluster.

In total, we can conclude that the GC star cluster is a uniform, well phase-mixed system. This is in good agreement with the age estimate for the cluster. Maness et al. (\cite{maness2007}) find constant star formation for $>$12~Gyr, meaning that most stars are older than $\approx10^9$~yr, the cluster's two-body relaxation time (Alexander~\cite{alexander2005}). In the following, we will discuss how our findings might help to constrain the dynamical history of the nuclear cluster. The most important question here is: if the GC cluster experienced the infall of another stellar system (e.g., a small star cluster), would we be able to detect corresponding kinematic traces in our data?

This discussion ties in with the debate on the origin of the young ($\approx$6~Myr) early-type stars located in two disks centered on Sgr~A*, for which two mechanisms have been proposed: in-situ star formation (e.g. Levin \& Beloborodov \cite{levin2003}; Goodman \cite{goodman2003}) or infall of a star cluster (e.g. Gerhard \cite{gerhard2001}; McMillan \& Portegies Zwart \cite{mcmillan2003}). In their analysis, Paumard et al. (\cite{paumard2006}) conclude that the initial mass of an infalling cluster is limited to $\approx$17000~$M_{\odot}$ for the more massive disk. In contrast, the inpiraling-cluster-scenario requires initial cluster masses $>10^5~M_{\odot}$. From this, Paumard et al. (\cite{paumard2006}) conclude that the inpiraling-cluster-scenario is highly unlikely.

On the one hand, the limited number of data points (in phase space) sets a \emph{lower limit} on the numbers of stars involved in or affected by such an event. Events involving only a small fraction of the stellar population would be masked by Poisson noise and residual methodological uncertainties. For somewhat larger events, the cluster returns to a relaxed, (quasi) equilibrium state within about one two-body relaxation time; this time is about 1~Gyr (Alexander~\cite{alexander2005}). However, phase-mixing (e.g. Binney \& Tremaine \cite{binney1987}) might erase the phase-space signature of any infall event already within few dynamical times ($t_{\rm dyn}\approx10^5$~yr for the GC cluster; Alexander~\cite{alexander2005})

On the other hand, violent relaxation (Lynden-Bell \cite{lyndenbell1967}; for a recent review, see Bindoni \& Secco \cite{bindoni2008} [and references therein]) sets an \emph{upper limit} on the amount of a dynamical distortion we could detect. In case of the infall of a massive object (in the order of the mass of the nuclear cluster), the cluster returns to a relaxed, (quasi) equilibrium state within about one dynamical time. This means that we would miss infalls of massive objects longer ago than few $10^5$ years anyway.

To summarize this discussion: we do not detect obvious kinematic signatures of an infall event. This indicates that there has been no major distortion of the GC cluster at least within the last few $10^5$ years. At this stage we are not able to quantify our statement further. We do expect, however, that a more detailed analysis, possibly in combination with dynamical modelling, will make the phase-space distribution analysis a valuable tool for characterising the kinematics of the GC cluster.

\subsection{Statistical parallax of the Galactic centre}

The availability of 3-dimensional velocity vectors for several hundred stars allows the computation of the distance to the Galactic centre $R_0$ using the statistical parallax. If the stellar velocities are distributed isotropically, then the three velocity dispersions $\sigma_{x}, \sigma_{y}, \sigma_{z}$ are equal. As $\sigma_{x}, \sigma_{y}$ are measured in angular units (mas/yr) whereas $\sigma_{z}$ is measured in physical units (km/s), the distance scale can be derived directly.

As discussed in the previous sections, in case of the GC cluster the global isotropy is broken by rotation. For the three velocity dispersions in $l$, $b$, and $z$ of our 664 3D velocity stars we find the values
\\

$\sigma_l = 2.928 \pm 0.082 {\rm \ mas/yr}$

$\sigma_b = 2.531 \pm 0.071 {\rm \ mas/yr}$

\begin{figure}
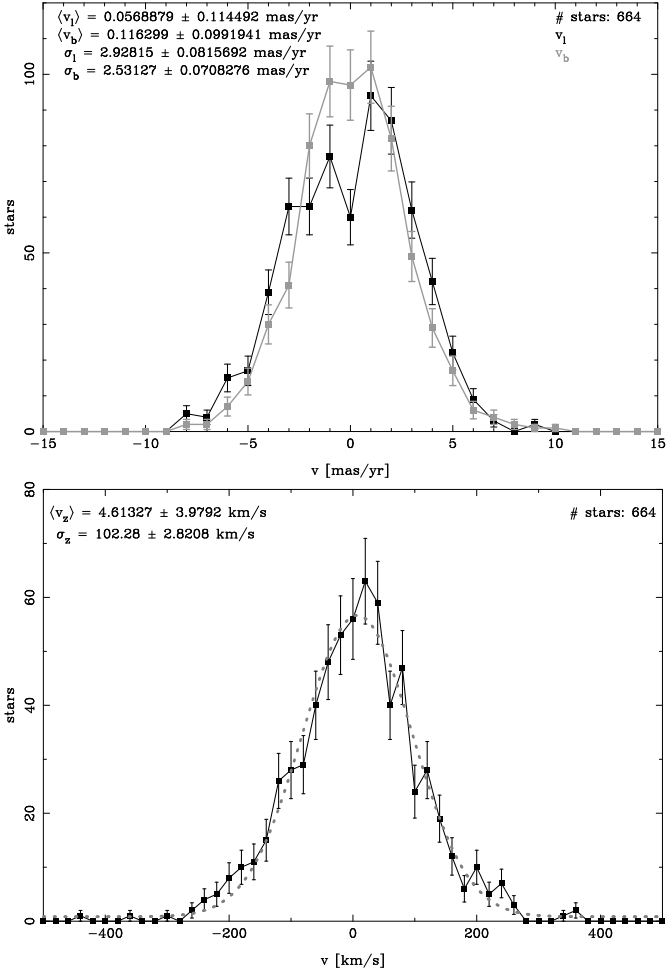
 

\includegraphics[height=8.8cm,angle=-90]{0191f18a.eps}

\includegraphics[height=8.8cm,angle=-90]{0191f18b.eps}

\caption{Histograms of proper motions in $l$ and $b$ ({\it top panel}) and of radial velocities ({\it bottom panel}) including all stars with measured 3D velocities. The dotted curve in the bottom panel corresponds to the best-fitting Gaussian profile. Error bars mark the Poisson errors. In both plots the respective average velocities $\langle \vel_{l, b, z}\rangle$ and velocity dispersions $\sigma_{l, b, z}$ are given.}

\label{dispersions3D}

\end{figure}

$\sigma_z = 102.3 \pm 2.8 {\rm \ km/s} ~.$
\\
\\
The respective distributions are shown in Fig. \ref{dispersions3D}. Our value for $\sigma_z$ is in excellent agreement with the value of 100.9$\pm$7.7~km/s found by Figer et al. (\cite{figer2003}) who analysed a smaller sample of 85 CO absorption line stars (recently, Zhu et al. (\cite{zhu2008}) confirmed this result).

In order to calculate a consistent distance estimate, we need to take into account two effects:

(1) The cluster is rotating in the $b=0$ plane. We therefore use the assumption that the nuclear cluster is an axisymmetric elliptical system with respect to the $b$ axis. This means that $\sigma_l$ and $\sigma_z$ (but not $\sigma_b$ and $\sigma_z$) can be compared in order to calculate the statistical parallax.

(2) We need to include projection effects which affect the impact of rotation. Whereas the observed $\sigma_l$ is given by Eq.~\ref{eq_jeans_sigmal}, the observed $\sigma_z$ is given by the relation

\begin{equation}
\Sigma (r) \sigma_z^2 (r) = 2\int_r^{\infty}\left(\sigma_l^2(R) + \tilde{\vel}(R)^2\frac{r^2}{R^2}\right)\frac{\rho(R) R}{\sqrt{R^2 - r^2}}dR
\label{eq_jeans_sigmaz} ~.
\end{equation}

\noindent
This relation differs from Eq.~\ref{eq_jeans_sigmal} in the geometry factor multiplied with $\tilde{\vel}^2(R)$; especially, this factor implies that $\sigma_l>\sigma_z$ in projection for a finite FOV. The use of $\sigma_l(R)$ to calculate $\sigma_z(r)$ is due to the assumption of axisymmetry.

From a comparison of Eq.~\ref{eq_jeans_sigmal} and Eq.~\ref{eq_jeans_sigmaz} we obtain a projection correction factor

\begin{equation}
\hspace{15mm} k = \frac{\sigma_l(r\rightarrow{\rm FOV})}{\sigma_z(r\rightarrow{\rm FOV})} \ge 1
\label{eq_parallax_projcorr} ~.
\end{equation}

\noindent
Here $\sigma_{l,z}(r\rightarrow{\rm FOV})$ are the projected velocity dispersions of the entire sample calculated in identical units (either mas/yr or km/s). The case $k=1$ corresponds to an infinite FOV; for FOVs covering only the central part of the cluster, $k>1$. From this relation we obtain the effective velocity dispersion in $z$,

\begin{equation}
\hspace{15mm} \sigma_z^{\prime} = k\sigma_z = \sigma_l
\label{eq_parallax_sigmazeff}
\end{equation}

\noindent
where the second equality is given by the assumption of axisymmetry. From our calculations we derive $k=1.095$, leading to
\\

$\sigma_z^{\prime} = 112.0 \pm 3.1 {\rm \ km/s} ~.$
\\

\noindent
Comparison of $\sigma_l$ and $\sigma_z^{\prime}$ leads to $R_0=8.07\pm0.32$~kpc.

The error quoted above is the \emph{statistical} uncertainty. Additionally, \emph{systematic} errors are introduced by selecting specific rotation profiles, (an)isotropic dispersion profiles, stellar density distributions, and other kinematic parametrizations (see sect.~5.3). From testing a variety of model parametrizations, we find a systematic error of $0.13$~kpc.

All in all, we derive a statistical parallax for the nuclear cluster of
\\

$R_0 = 8.07 \pm 0.32_{\rm stat} \pm 0.13_{\rm sys} {\rm \ kpc} ~.$
\\

\noindent
This result is in full agreement with those obtained from the observations of Keplerian stellar orbits around Sgr~A* (e.g. Eisenhauer et al. \cite{eisenhauer2005}; Lu et al. \cite{lu2006}). The most recent values are given by Gillessen et al. (\cite{gillessen2008}) who find $R_0=8.14\pm0.15_{\rm stat}\pm0.32_{\rm sys}{\rm \ kpc}$ and by Ghez et al. (\cite{ghez2008}) who find $R_0=8.0\pm0.6_{\rm stat+sys}{\rm \ kpc}$.

Our result also agrees with distance values obtained by earlier statistical parallax measurements, which were $7.9\pm0.9_{\rm stat}$~kpc (Genzel et al. \cite{genzel2000}) and, more recently, $7.1\pm0.7_{\rm stat}$~kpc (Eisenhauer et al. \cite{eisenhauer2003a}). In the aforementioned experiments the systematic difference between the dispersions in $l$ and $b$ directions was masked by larger statistical uncertainties; therefore, the influence of cluster rotation was not recognized.

Other experiments are based on precision stellar photometry (e.g. Paczynski \& Stanek \cite{paczynski1998}; McNamara et al. \cite{mcnamara2000}). Recent results have been $R_0=7.52\pm0.10_{\rm stat}\pm0.35_{\rm sys}{\rm \ kpc}$ by Nishiyama et al. (\cite{nishiyama2006}) and $R_0=7.94\pm0.37_{\rm stat}\pm0.26_{\rm sys}{\rm \ kpc}$ by Groenewegen et al. (\cite{groenewegen2008}). All these results are in full agreement with our value.

Another method used to derive $R_0$ is the spatial distribution of globular clusters in the Milky Way. In a recent analysis, Bica et al. (\cite{bica2006}) examined a sample of 116 clusters and derived $R_0=7.2\pm0.3_{\rm stat}$~kpc. Within errors, this is only marginally in agreement with our findings; but when taking into account that Bica et al. (\cite{bica2006}) do not provide an estimate for the systematic uncertainty of their method, this deviation is not significant either. 

In total, we can conclude the following: (1) The statistical parallax is a valuable independent method for deriving $R_0$. (2) Our value and recent measurements based on different approaches (stellar orbits, stellar photometry, distribution of globular clusters) are in good agreement.

\subsection{Acceleration upper limits}

The strong influence of Sgr~A* allows the description of the innermost part (few arcsec) of the cluster as a system of massless test particles moving around a point mass on Keplerian orbits. Indeed, orbits located in the innermost $\approx0.5$'' -- in the so-called ``S-star'' group -- have been observed now for several years without detecting any significant deviation from a point mass potential (Eisenhauer et al. \cite{eisenhauer2005}; Lu et al. \cite{lu2006}; Gillessen et al. \cite{gillessen2008}; Ghez et al. \cite{ghez2008}). When using the very accurate proper motions obtained from the small scale images (typical uncertainties $\approx$4~km/s, see Fig.~\ref{velerr}), it is possible to detect (or exclude) accelerations in stellar motion as far out as several arcseconds in projected distance.

In this section and Section~5.7 we analyse all available stars \emph{regardless of their spectral type}. We can do this because we treat stars individually. Global statistical properties like isotropy, which are affected by the non-relaxed early-type population, do not influence this analysis.

\begin{figure} 

\includegraphics[height=8.8cm,angle=-90]{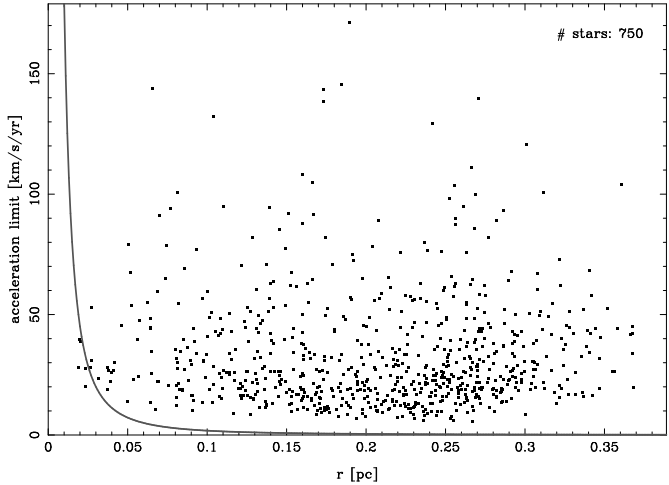}

\caption{Acceleration upper limits (99\% confidence level) vs. projected distance $r$ from Sgr~A*. Dots mark measured values. The continuous line corresponds to the acceleration a star would experience if its physical distance from Sgr~A* would equal its projected distance. The distance scale is 1~pc~$\equiv$~25.8''. This sample includes 750 of 755 analysed stars. Stars below the line have a physical distance that is necessarily larger than the projected distance.}

\label{acc_vs_r}

\end{figure}

In order to obtain acceleration limits we analysed the proper motions of stars in the small scale fields within $\pm$7'' in R.A. and DEC from Sgr~A*. All star positions were transformed into coordinates radial ($q_{\parallel}$) and tangential ($q_{\perp}$) to their average position vector. In analogy to the determination of proper motions we fit the star positions ${\bf q}=(q_{\parallel},q_{\perp})$ vs. time $t$ as parabolas of the form

\begin{equation}
\hspace{10mm} {\bf q}(t) = {\bf u} t^2 + {\bf v} t + {\bf w} ~.
\label{eq_acceleration}
\end{equation}

\noindent
Obviously, this approach delivers the (constant) stellar accelerations $\bf a$ (via ${\bf a} = 2{\bf u}$); {\bf v} and $\bf w$ correspond to velocities and positions at $t=0$ respectively.

Whether a given star shows a significant acceleration depends on the goodness-of-fit of the two physically realistic models, which are (1) a linear proper motion, and (2) an accelerated parabolic motion pointing towards Sgr~A* (i.e. a significantly non-zero value for $a_{\parallel}$). For both models the respective reduced $\chi^2$ (hereafter $\chi^{2}_{\rm lin}$ for the linear, $\chi^{2}_{\rm acc}$ for the accelerated case) is computed.

In order to decide if the difference between the two models is significant we make use of the fact that the quantity

\begin{equation}
\hspace{10mm} f = \frac{\chi^{2}_{\rm lin}}{\chi^{2}_{\rm acc}}
\label{eq_ftest}
\end{equation}

\noindent
follows an F-distribution and can thus be examined using an F-test (e.g. M\"uller \cite{mueller1975}; Lehn \& Wegmann \cite{lehn1982}). For a given significance level $S\in [0,1]$ and the case of a one-tailed test (e.g. Snedecor \& Cochran \cite{snedecor1989}) the difference is considered to be significant if $f > F_{m, n, 1-s}$; here $m, n$ are the respective degrees of freedom, $s = 1-S$ is the false alarm probability. For the typical case $m = n+1 = 35$ we find $F_{35, 34, 0.99} = 2.25$ for $S = 0.99$. This means that a 99\% confidence detection of non-linear motion requires $f > 2.25$.

Our analysis included a total of 755 stars. For five of them significant (using a 99\% confidence limit) accelerations were detected. Three stars were S-stars with known orbits. Two more were false positives whose centroids were systematically displaced towards bright neighbouring sources. Due to the high crowding in the innermost arcseconds of the field most of the S-stars known to follow Keplerian orbits had to be excluded from this automated analysis and therefore do not contribute.

For the remaining 750 sources, acceleration upper limits are given. Upper limits with a 99\% confidence (2.58$\sigma$) are defined as

\begin{equation}
\hspace{15mm} \bar{a}_{\parallel} = |a_{\parallel}| + 2.58\times\delta a_{\parallel} ~.
\label{eq_acclimit}
\end{equation}

\noindent
Here $|a_{\parallel}|$ is the amount of radial acceleration computed according to Eq.~\ref{eq_acceleration} and $\delta a_{\parallel}$ its 1$\sigma$ error\footnote{Even if a star does not show significant acceleration, a parabolic fit finds a (usually non-zero) result for the 2nd-order term and a corresponding fit error. These values we use here.}. Fig.~\ref{acc_vs_r} shows the resulting limits vs. projected distances. For comparison, we show the acceleration a star would experience if its projected distance were equal to its physical distance. A few limits (5 out of 750) fall below the theoretical line; in these cases, the physical distances need to be larger than the projected distances.

%\begin{figure} 
%\includegraphics[height=8.8cm,angle=-90]{acclimitminz.eps}
%\caption{Minimum distance from the plane of Sgr~A*, $|z|_{\rm min}$, vs. projected radius $r$ for five stars with $R_{\rm min} > r$. The values are based on 2.58$\sigma$ (99\% confidence level) acceleration upper limits. The distance scale is 1~pc~$\equiv$~25.8''. Only stars with $r<0.025$~pc are sufficiently constrained; see Fig.~\ref{acc_vs_r}.}
%\label{accprofile}
%\end{figure}

\begin{table} 

\centering

\caption{Minimum distance from the plane of Sgr~A*, $|z|_{\rm min}$, vs. projected radius $r$ for five stars with $R_{\rm min} > r$. The values are based on 2.58$\sigma$ (99\% confidence level) acceleration upper limits. Only stars with $r<0.024$~pc are sufficiently constrained; see Fig.~\ref{acc_vs_r}.}

\begin{tabular}{l c c}
\hline\hline
Star ID & r [mpc] & $|z|_{\rm min}$ [mpc] \\
\hline
823 & 19.2 & 12.9 \\
832 & 23.2 & 8.6 \\
1333 & 20.1 & 6.4 \\
1760 & 19.4 & 7.3 \\  
3439 & 23.5 & 14.5 \\  
\hline
\end{tabular}

\label{accprofile}
\end{table}

An acceleration upper limit constrains the minimum physical distance $R_{\rm min}$ of a star. We can compare this number with the star's sky-projected 2D distance $r$. Starting from the amount of sky-projected acceleration

\begin{equation}
\hspace{15mm} a_{\rm 2D} = \frac{GM}{R^3}r
\label{eq_accel2D}
\end{equation}

\noindent
and replacing $a_{\rm 2D}$ by $\bar{a}_{\parallel}$ we obtain

\begin{equation}
\hspace{15mm} R > \left(\frac{GM}{\bar{a}_{\parallel}}r\right)^{1/3} \equiv R_{\rm min}
\label{eq_accradius}
\end{equation}

\noindent
where $R$ is the physical 3D distance from Sgr~A*. If $R_{\rm min} > r$ for a star, we can derive its minimum distance from the plane of Sgr~A* along the line of sight

\begin{equation}
\hspace{15mm} |z|_{\rm min} = \sqrt{R_{\rm min}^2 - r^2} ~.
\label{eq_accminz}
\end{equation}

\noindent
In our case, we have five stars with $R_{\rm min} > r$ for which we can derive the corresponding $|z|_{\rm min}$; all have $r<0.024$~pc. The resulting distribution is presented in Table~\ref{accprofile}; the largest distance we find is $|z|_{\rm min}\approx0.015$~pc.

Given the good accuracies of the stellar proper motions (few km/s), the large values for the acceleration limits (tens of km/s/yr; see Fig.~\ref{acc_vs_r}) might not be obvious. However, this is given by the fact that both velocities and accelerations are derived from position measurements. Whereas the statistical error of a \emph{velocity} measurement scales like

\begin{equation}
\hspace{15mm} \delta \vel \propto \delta q \times t^{-1} ,
\label{eq_velerr}
\end{equation}

\noindent
the statistical error of an \emph{acceleration} measurement scales like

\begin{equation}
\hspace{15mm} \delta a \propto \delta q \times t^{-2}
\label{eq_accerr}
\end{equation}

\noindent
with $\delta q$ being the typical position measurement uncertainty, $t$ being the total time covered by data. From this, we see that on the one hand a timeline of five years is sufficient for precision proper motion measurements. On the other hand, the same timeline of five years is not yet sufficient to constrain the $z$ coordinate of more than a handful of stars. However, as $\delta a \propto t^{-2}$, already a moderate extension of the timeline leads to a substantial improvement.

From this we can draw important conclusions affecting future analyses and observing strategies. (1) Future works should include earlier data. For the GC, we have data at hand starting as early as 1992, taken with the SHARP~I speckle imaging camera at the ESO-NTT (e.g. Genzel et al. \cite{genzel1996}). For the case of the S-stars orbiting Sgr~A*, this dataset has been used with great success (Sch\"odel et al. \cite{schoedel2003}; Gillessen et al. \cite{gillessen2008}). (2) Continued NACO observation of the nuclear cluster will provide valuable new insights. Therefore the monitoring of the cluster should be continued.

\subsection{The star group IRS13E}

An object of special interest is the star group IRS13E, located 3'' west and 1.5'' south of Sgr~A*. This object consists of three bright (H$\approx$13) main components concentrated within a region of about 0.2'' radius. They surround fainter objects which are probably blends of several point sources. Paumard et al. (\cite{paumard2006}) found a significant stellar density excess in the immediate vicinity (0.7'') of the three main stars and identified the IRS13E group as a star cluster. Based on stability arguments with respect to the tidal field of Sgr~A*, the possibility that IRS13E hosts an intermediate-mass black hole was previously discussed by Maillard et al. (\cite{maillard2004}).

In order to examine this scenario in more detail, we tested whether stars within a radius of 0.7'' are kinematically connected with the three central sources. We extracted proper motions for the three main components and an additional 17 stars with magnitudes down to H$\approx$19.5 (for a similar analysis using the proper motions of another set of stars, see Sch\"odel, Eckart \& Iserlohe \cite{schoedel2005}). In the following, we call the three main stars ``set A'' and the 17 field stars ``set B''. As the target area is too crowded to be fully covered by the automatic procedures described in Section 3, we extracted image positions manually. We applied the PSF fitting routine {\it StarFinder} by Diolaiti et al. (\cite{diolaiti2000}) and checked the inter-epoch source identifications by eye. Our analysis included seven very good H and K-band small scale images obtained between 2002 and 2007.

\begin{figure}
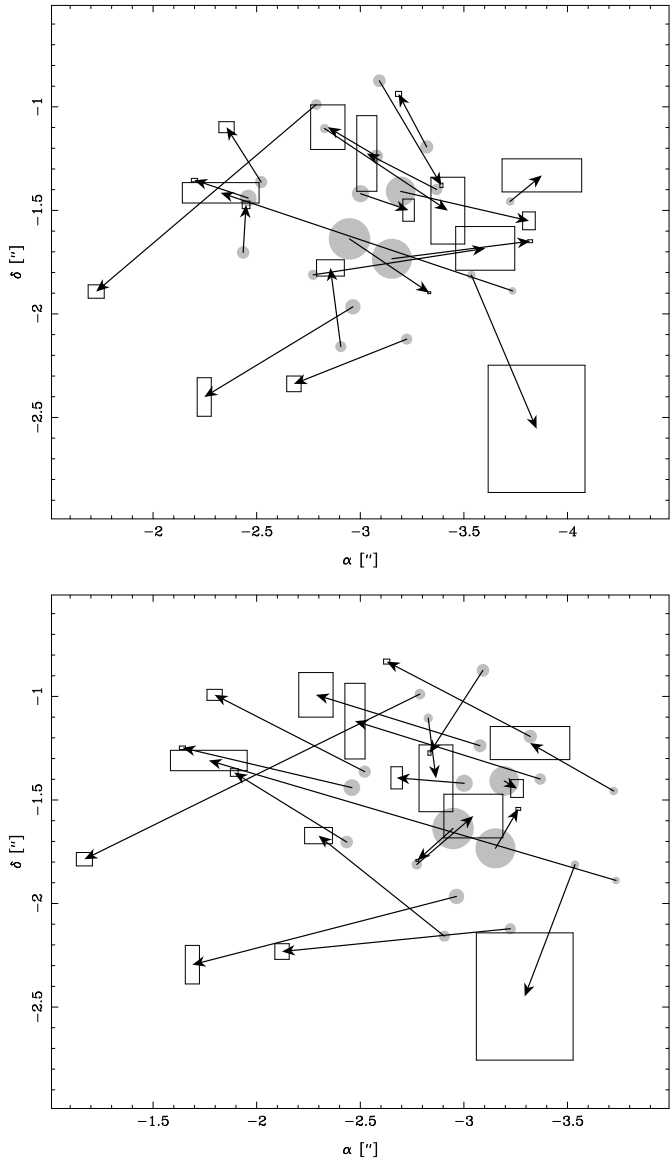
 

\includegraphics[height=8.8cm,angle=-90]{0191f20a.eps}

\vspace{3mm}

\includegraphics[height=8.8cm,angle=-90]{0191f20b.eps}

\caption{Proper motion maps for the IRS13E group. Filled circles are stars, circle diameters are proportional to the 0.25th power of H-band fluxes. Coordinates are $\alpha$, $\delta$ in arcsec relative to Sgr~A*; please note the shift in $\alpha$. Arrows are proper motions, 1 unit in length corresponds to 10~mas/yr. Boxes around arrow heads indicate the 1-$\sigma$ velocity uncertainties. \emph{Top panel}: Proper motions co-moving with the GC cluster as defined in section 3. \emph{Bottom panel}: Proper motions co-moving with the central three brightest stars. A streaming motion of the three brightest stars relative to a foreground/background population can be seen. Apparently, only a small fraction, if any, of the examined stars are bound to the IRS13E group.}

\label{irs13}

\end{figure}

The main results are summarized in Fig.~\ref{irs13}. The top panel shows all stellar proper motions with respect to the standard astrometric reference frame tied to the GC cluster, i.e. co-moving with the GC cluster (see Section 3). The bottom panel of this figure shows the same proper motions in a reference frame co-moving with set A, using the average motion of set A as the zero-point. In the latter case, almost all stars show a motion directed from west to east, as if IRS13E were moving through a separate, non-co-moving foreground/background population. In $\alpha$ (or $x$) direction, the average motion (co-moving with set A) of set B is $\langle \vel_x\rangle = 7.1\pm1.3$~mas/yr, wheras the velocity dispersion is $\sigma_x = 5.3\pm0.9$~mas/yr. In $\delta$ (or $y$), we find $\langle \vel_y\rangle = 0.6\pm0.9$~mas/yr, $\sigma_y = 3.7\pm0.7$~mas/yr. These numbers show a clear streaming of the three main stars with respect to the 17 field stars.

For their analysis, Paumard et al. (\cite{paumard2006} [see especially their Fig.~9]) used stellar number counts in the IRS13E region down to magnitudes H$=$20.4. They compared the surface densities inside ($\Sigma_<$) and outside ($\Sigma_>$) a projected radius of $p=0.68$'' from the centre of set A. Counting all H$<$20.4 stars they found
\\

$\Sigma_< = 31.7 \pm 4.7 \ \rm arcsec^{-2}$

$\Sigma_> = 13.1 \pm 1.2 \ \rm arcsec^{-2} ~.$
\\

\noindent
This corresponds to a 4.3$\sigma$ excess of stars within $p$. Using a more conservative limit of H$<$19.4, the resulting densities are
\\

$\Sigma_< = 17.9 \pm 3.5 \ \rm arcsec^{-2}$

$\Sigma_> =~~7.9 \pm 0.9 \ \rm arcsec^{-2}$
\\

\noindent
corresponding to a 2.8$\sigma$ excess of stars within $p$. All errors given for numbers and densities are Poisson errors. The total number of H$<$19.4 stars located within $p$ is 26. This magnitude cut corresponds (within the errors) to the magnitude limit of our set B.

We used the kinematic information obtained for the set B stars to follow up on the surface density analysis by Paumard et al. (\cite{paumard2006}). We recalculated the surface density of H$<$19.4 stars, but excluded stars which are too fast with respect to set A. The star selection was done in two steps. First, we computed the 2D velocity dispersion of the three set A stars. This dispersion was $\sigma_{\rm A} = 1.9 \pm 0.8 \ \rm mas/yr$.

Then we computed the bias-corrected 2D speeds $\vel_{\rm 2D}$ of the set B stars relative to set A (reference frame co-moving with set A). We identified all set B stars (a) located within $p$ and (b) showing $\vel_{\rm 2D}>3\sigma_{\rm A}$. For these stars we assumed that they cannot be physically connected to the set A stars.

We found nine stars satisfying criteria (a) and (b). Excluding them from the sample of Paumard et al. (\cite{paumard2006}) reduces the number of H$<$19.4 stars located within $p$ from 26 to 17. Recalculating the surface density leads to
\\

$\Sigma_< = 11.7 \pm 2.8 \ \rm arcsec^{-2} ~.$
\\

\noindent
The difference of densities inside and outside $p$ then becomes
\\

$\Delta\Sigma = \Sigma_< - \Sigma_> = 3.8 \pm 3.0 \ \rm arcsec^{-2}$
\\

\noindent
corresponding to a significance of 1.3$\sigma$ for a deviation from zero. We therefore do not see a significant excess of H$<$19.4 stars within $p$.

We can conclude that our kinematic analysis seriously weakens the scenario proposing that IRS13E is the core of a substantial star cluster.

\section{Conclusions}

In this article we analysed and discussed the kinematic properties of the Galactic centre CO absorption line star cluster. This work is based on adaptive optics assisted diffraction-limited near-infrared imaging and integral-field spectroscopy. We collected proper motions for 5445 stars, 3D velocities for 664 stars, and acceleration limits for 750 stars. Our analysis led to the following main results:

\begin{enumerate}

\item The cluster shows a global rotation in the sense of general Galactic rotation.

\item The stellar 3D speed distributions can be locally approximated by Maxwellians. This confirms the relaxed nature of the CO star cluster. 

\item We find one high-velocity star with a 3D speed $\vel_{\rm 3D}=810\pm9$~km/s which might be unbound. It might have been ejected from the cluster by three-body interactions with SgrA*.

\item We obtain a deprojected 3D description of the cluster kinematics. We fit the observed velocity dispersion profiles $\sigma_{l,b}(r)$ and the rotation curve $\langle \vel_z \rangle (l)$. The data are described well by assuming a $\rho(R)\propto R^{-2}$ sphere density profile and global rotation. From the model solution, we extract the cluster's mass profile out to $M_{*}\simeq2.3\times10^7~M_{\odot}$ at $R\simeq4$~pc.

\item The two-point correlation function of the stellar 4D phase-space positions agrees with that of a uniform isotropic rotator within 4\%. We find no obvious indication for phase-space subtructure like star streams. From this we conclude qualitatively that there has been no major distortion of the GC cluster within the last few $10^5$ years.

\item Using the 3D velocity dispersion, we derive an improved statistical parallax to the GC of $R_0=8.07\pm0.32_{\rm stat}\pm0.13_{\rm sys}$~kpc. This result is in good agreement with the values obtained by stellar orbit, stellar photometry, and globular cluster distribution studies.

\item For stars located within the innermost few (projected) arcsec, we calculate limits on accelerations in the plane of sky. We use these limits to constrain the stars' minimum line-of-sight distances from the plane of Sgr~A*. We find non-trivial results for 5 out of 750 stars and conclude that already a moderate extension of the observation timeline can increase this number substantially.

\item The star group IRS13E does not co-move with the $H<19.4$ stars in its 0.7'' vicinity. When excluding stars which are too fast to be part of the IRS13E system, there is no sign for a significant star concentration. This seriously weakens the case for IRS13E being the core of a substantial star cluster.

\end{enumerate}

In summary, our analysis has improved substantially our knowledge regarding the kinematic properties of the GC star cluster. The next step will be to feed our extensive data set into a full-scale dynamical model. We plan to make use of the recently developed {\it NMAGIC} code (De Lorenzi et al. \cite{delorenzi2007}) in order to finalize the physical description of the old stellar population in the central parsec of our Milky Way.

\begin{acknowledgements}

Special thanks to Nancy Ageorges (ESO, MPE) for helpful discussions of the instrumental geometric distortion and registration of NACO images. We are grateful to Tal Alexander and Hagai Perets (both: Weizmann Institute of Science, Rehovot, Israel) for enlightening discussions of stellar dynamics. Many thanks to Nicolas Bouch\'e (MPE) for his description of the properties of two-point correlation functions, and to Rainer Sch\"odel (IAA Grenada) for helpful discussions of the GC cluster mass profile. We are grateful to Hauke Engel, Eva Noyola, and Richard Davies (all: MPE) for helpful discussions of the properties of the Jeans equation. S.T. would like to thank Markus Schmaus (LMU Munich, Dept. of Mathematics) for pointing out the concept of stochastic optimization. And last but not least we are grateful to the referee, James Binney, whose comments helped to improve the quality of this article.

\end{acknowledgements}

\appendix

\section{Image distortion parameters}

We extracted the distortion parameters from our imaging data by executing the following steps:

\begin{enumerate}

\item Combination of all individual frames to be mosaicked via simple shift-and-add (SSA) with integer-pixel accuracy.

\item Constructing a list of many ($\approx$200) good (meaning bright, but unsaturated stars well separated from neighbouring sources) reference stars distributed over the entire FOV. For source selection, the SSA image is used.
 
\item Re-identification of all reference stars located in the FOV of each individual image, followed by determining their detector positions.

\item Computation of all pairwise star--star separations in each image. This calculation results in a net of baselines for each image. Baselines present in more than one image are subject to inter-image comparison. 

\item  Modelling the distortion correction.

\end{enumerate}

\noindent
The three model parameters ($\beta$, $x_C$, $y_C$) were fit by a $\chi^2$ minimization. Geometric distortion implies that the detector plate scale is a function of the detector position; thus the length of a given baseline (in units of pixels) depends on its location on the detector. The optimum parameter set is found by iteratively comparing all baselines in all images shifted on sky, applying the temporary parameters to the reference star detector coordinates, and checking for the improvement.

In case of the \emph{large scale} (27~mas/pixel) images we executed parameter fits in a straight forward manner. We made use of the analytic fit engine {\it FindMinimum} implemented in the software package {\it Mathematica}.

For the \emph{small scale} (13~mas/pixel) data sets this procedure was not applicable. Due to the less significant distortion and the smaller number of reference stars available ($\approx$100), the analytic fit algorithm usually did not converge towards a reliable result. Thus we constructed a stochastic minimization algorithm which searches the parameter space iteratively using the following scheme:

\begin{enumerate}

\item Compute the value of the cost function (i.e. the function to be minimized) at the actual position in parameter space.

\item Select a second position in parameter space and compute the value of the cost function at that position.

\item If the value of the cost function at the new position is smaller than the actual one: move there. Otherwise: stay at present position.

\item Repeat steps 1--3 until a fixed number of iterations is completed.

\end{enumerate}

\noindent
Starting from a given initial position in parameter space, in iteration $n$ for each parameter $p$ a new value (step 2) is computed as

\begin{equation}
\hspace{7mm} p_{n+1} = p_n +  s(n) \cdot (1/z_{[0,1]} - 1) \cdot \epsilon (0.5-z_{[0,1]}) 
\label{eq_nminimze}
\end{equation}

\noindent
with

$$
s(n) = s_0 \cdot 10^{-n/N} ~.
$$

\noindent
Here $N$ is the maximum number of iterations, $z_{[0,1]}$ a random number in the range $[0,1]$, $\epsilon (x)$ the sign function returning $-1$ or $+1$ depending on the sign of $x$, and $s_0$ the initial step size.

This definition assures that (1) the algorithm cannot be easily trapped in a local minimum, as even extreme search radii are occasionally tested, and (2) the vicinity of the best-so-far-solution found at the end of the search time is explored with reasonable accuracy, as the average search radius decreases exponentially with time (thus increasing the ``selection pressure'' on the algorithm). The idea for this definition was taken from the concept of Simulated Annealing introduced by Kirkpatrick et al. (\cite{kirkpatrick1983}).

\section{Accuracy of radio coordinates}

Absolute radio coordinates and infrared image coordinates are tied via nine SiO maser stars in the field of view. These maser stars show finite statistical errors in both radio and NIR image positions. The uncertainties limit the accuracy of the global coordinates.

We examined the influence of these uncertainties on the transformation accuracy using a Monte-Carlo test. We executed $10^5$ coordinate transformations, each time using sets of positions with random displacements. The displacements followed Gaussian distributions according to the individual statistical errors.

By sampling a coordinate grid with the typical FOV size (positions $\pm$20'' from Sgr~A*) we mapped the transformation uncertainty as a function of position. The results are shown in Fig. \ref{trafo}. The contours mirror the geometry of the alignment of the reference stars in the plane of the sky (see also Fig. \ref{field}); the errors vary in the range 1.2~...~5~mas.

Of particular interest is the accuracy of the global position (0,0) which corresponds to the location of Sgr~A*, the dynamical centre of the GC star cluster. Here the errors were
\\

$ \delta {\rm R.A.} = 1.26 {\rm\ mas} $

$ \delta {\rm DEC} = 1.20 {\rm\ mas} ~.$
\\

\noindent
These errors are uncertainties in \emph{absolute} positions. Our kinematic analysis is based on time-resolved \emph{relative} positions. Therefore the proper motion accuracies are much better than suggested by the errors of the absolute positions.

\end{document}